\documentclass[preprint,12pt,authoryear]{elsarticle}
\usepackage{enumitem}
\usepackage{natbib}
\usepackage{booktabs}
\usepackage{multirow}
\usepackage{array}
\usepackage{url}
\usepackage{hyperref}
\usepackage{siunitx}
\usepackage{amssymb}
\usepackage{subcaption}
\usepackage{graphicx}
\usepackage{listings}
\usepackage{xcolor}  
\usepackage[utf8]{inputenc}
\usepackage{amsmath}

\journal{Computers \& Education}

\begin{document}

\begin{frontmatter}



\title{Are LLMs Ready for Computer Science Education? A Cross‑Domain, Cross‑Lingual and Cognitive-Level Evaluation Using Professional Certification Exams} 


\author[0]{Chen Gao}

\author[1]{Chi Liu\corref{cor1}}
\ead{chiliu@cityu.edu.mo}

\author[1]{Zhengquan Luo}

\author[1]{Dongfu Xiao}

\author[2]{Maiying Sui}

\author[3]{Sheng Shen}

\author[1]{Congcong Zhu}

\author[1]{Huajie Chen}

\author[1]{Xuhan Zuo}

\author[4]{Zongyuan Ge}

\author[1]{Tianqing Zhu}

\author[1]{Wanlei Zhou}

\author[0]{Xiaotong Han\corref{cor1}}
\ead{hanxiaotong2@gzzoc.com}

\cortext[cor1]{Xiaotong Han and Chi Liu are Corresponding authors}

\affiliation[0]{organization={Zhongshan Ophthalmic Center, Sun Yat-sen University, State Key Laboratory of Ophthalmology, Guangdong Provincial Key Laboratory of Ophthalmology and Visual Science, Guangdong Provincial Clinical Research Center for Ocular Diseases},
            city={Guangzhou},
            country={China}}
            
\affiliation[1]{organization={Faculty of Data Science, City University of Macau},
            addressline={Avenida Padre Tomás Pereira},
            city={Taipa},
            postcode={999078},
            state={Macao},
            country={China}}

\affiliation[2]{organization={Department of Accounting, Faculty of Business and Economics, The University of Melbourne},
            addressline={198 Berkley St},
            city={Carlton},
            postcode={3010},
            state={VIC},
            country={Australia}}
            
\affiliation[3]{organization={School of Information Technology, Faculty of Business and Hospitality, Torrens University Australia},
            addressline={196 Flinders St},
            postcode={3000},
            state={VIC},
            country={Australia}}
            
\affiliation[4]{organization={AIM for Health Lab, Faculty of IT, Monash University, Australia},
            addressline={25 Exhibition Walk},
            city={Clayton},
            postcode={3800},
            state={VIC},
            country={Australia}}

\begin{abstract}

Large language models (LLMs) are increasingly applied in computer science education for tasks such as tutoring, content generation, and code assessment. However, systematic evaluations aligned with formal curricula and certification standards remain limited. This study benchmarked four recent models, including GPT-5, DeepSeek-R1, Qwen-Plus, and Llama-3.3-70B-Instruct, using a dataset of 1,068 questions derived from six certification exams covering networking, office applications, and Java programming.

We evaluated performance across language (Chinese vs.\ English), cognitive levels based on Bloom's Taxonomy, domain knowledge, confidence--accuracy alignment, and robustness to input masking. Results showed that GPT-5 performed best on English-language certifications, while Qwen-Plus performed better in Chinese contexts. DeepSeek-R1 achieved the most balanced cross-lingual performance, whereas Llama-3.3 showed clear limitations in higher-order reasoning and robustness. All models performed worse on more complex tasks.

These findings provide empirical support for the integration of LLMs into computer science education and offer practical implications for curriculum design and assessment.

\end{abstract}

\begin{graphicalabstract}
\includegraphics[width=\textwidth]{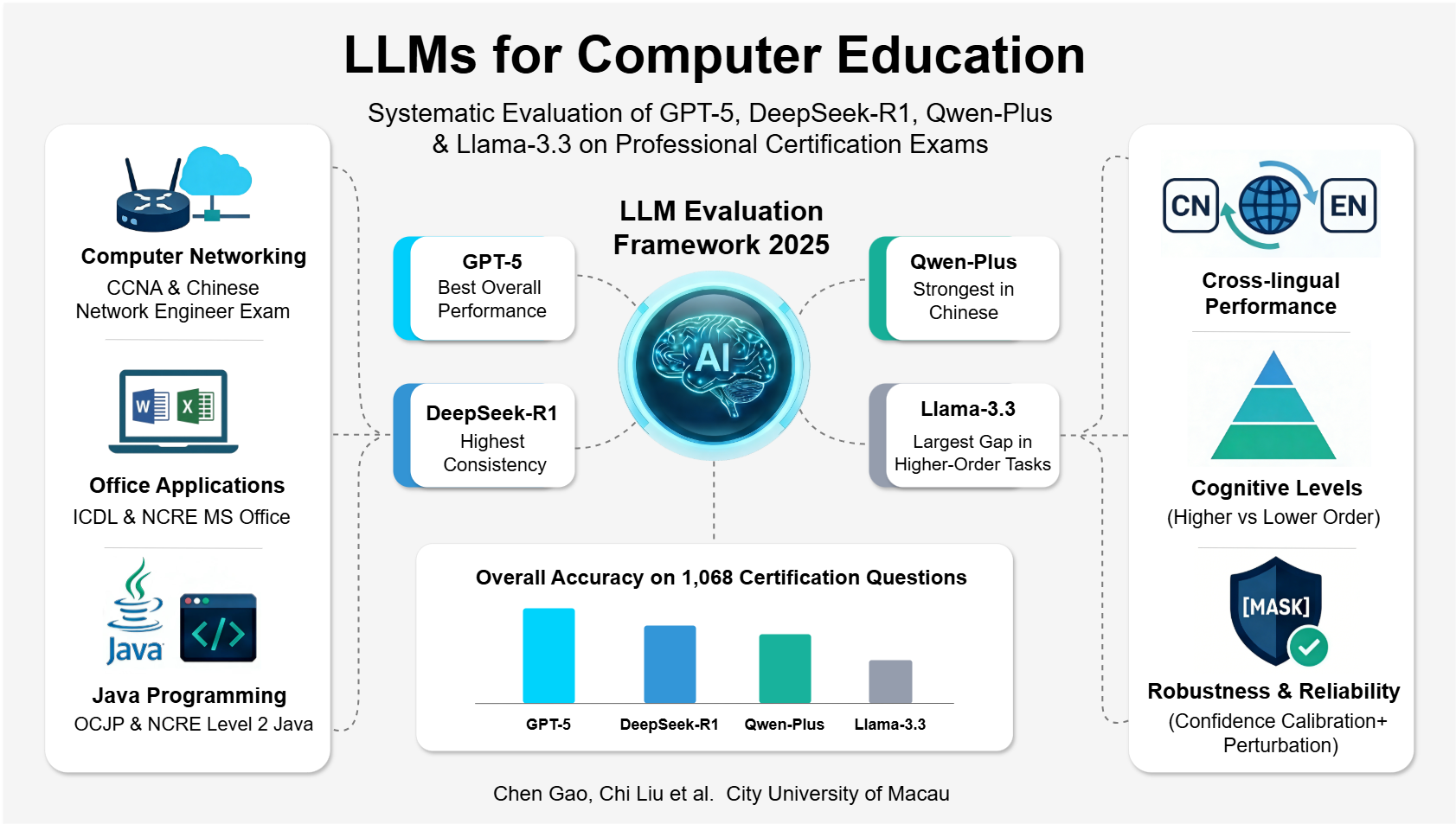} 
\end{graphicalabstract}

\begin{highlights}
\item Comprehensive benchmarking of four state-of-the-art LLMs (GPT-5, DeepSeek-R1, Qwen-Plus, Llama-3.3-70B-Instruct) on 1,068 questions from six international and Chinese computer science certification examinations.
\item GPT-5 achieves the highest accuracy on English/international certifications and exhibits superior robustness to input perturbations and higher-order reasoning tasks.
\item Qwen-Plus excels on Chinese domestic exams while DeepSeek-R1 provides the most balanced cross-lingual and cross-domain performance.
\item All models show substantial performance degradation on higher-order cognitive tasks (Bloom’s taxonomy) and under information masking, with Llama-3.3 being most affected—highlighting current limitations for advanced CS education applications.
\end{highlights}


\end{frontmatter}



\section{Introduction}
\label{sec:intro}

Currently, large language models (LLMs) have demonstrated superior performance in multiple fields such as natural language understanding, complex reasoning, and code generation \citep{brown2020language,wei2022chain,achiam2023gpt}.
These capabilities demonstrate their transformative potential in computer science education, enabling them to be used for tasks including automated tutoring, exercise generation, code quality assessment, and personalized feedback \citep{peng2023impact,xu2024large,wang2024large}.
Most previous studies have used general benchmarks (e.g., MMLU, GSM8K) or broad coding tasks (e.g., HumanEval) to evaluate the performance of large language models, while rigorous assessments for specific areas in computer science education remain relatively scarce \citep{xiao2025can}, and those tests are often poorly aligned with formal curriculum structures, hierarchical skill development, and standardized certification exams \citep{chang2024survey,liang2022holistic}.

Furthermore, some studies rely on limited assessment metrics, such as pass rates for discrete programming problems, rather than adopting a multidimensional framework based on recognized educational and industry standards. This limits our comprehensive understanding of the capabilities and limitations of LLMs in core sub-domains of computer science education, including computer networks, office software proficiency, and advanced programming. To address these shortcomings, this study systematically evaluated four advanced large-scale language models representing trends expected by the end of 2025: DeepSeek-R1, Qwen-Plus (a balanced variant of the Qwen3 series), Llama-3.3-70B-Instruct, and GPT-5. The selected models excelled in mathematical reasoning, code generation, and logical analysis—the cornerstones of computer science education.

The evaluation framework employed in the experiment comprised six interconnected and subject-specific modules based on established certification exams: computer networking (Cisco Certified Network Engineer [CCNA] and the China Network Engineer exam within the National Software Professional Technical Qualification System), office software proficiency (the Microsoft Office portion of the International Computer Driving License [ICDL] and the China National Computer Skills Examination [NCRE]), and Java programming (the Oracle Certified Professional Java SE Programmer [OCJP] certification and the NCRE Level 2 Java module).

By building the evaluation system around these established educational and industry benchmarks, our multidimensional framework supports a more systematic evaluation of the models—an evaluation that is both domain-specific and applicable to different languages. This is especially important in rigorous, high-stakes testing environments such as certification exams. In addition to using benchmarks to compare and analyze the strengths and limitations of the models in core sub-domains of computer science education, the findings also provided empirical evidence to guide how to responsibly integrate learning models into computer science instruction, assessment design, and learner support systems.

The study is guided by two research questions:

\textbf{RQ1:} Are there systematic differences in the performance of LLMs across different types of computer science education assessments? While these models typically perform well on general tasks, their results may fluctuate dramatically when faced with specific professional exams.
We propose the following three hypotheses:
\begin{enumerate}
    \item \textbf{Language Disparity}: Significant performance gaps exist between Chinese and English tasks, potentially affecting the models' efficacy across multilingual pedagogical environments.
    \item \textbf{Domain Sensitivity}: LLM capabilities vary substantially across distinct computer science subdomains, likely reflecting disparities or biases in the distribution of their pre-training data.
    \item \textbf{Cognitive demand}: The cognitive demands of a task—ranging from lower-order recall to higher-order analysis and application—significantly impact model performance, with proficiency expected to decline as cognitive complexity increases.
\end{enumerate}

Clarifying these differences will inform more targeted selections of artificial intelligence tools in practical teaching scenarios.

\textbf{RQ2:} How robust and reliable are the outputs of LLMs in computer science education? The reliability of artificial intelligence depends on the robustness of its performance. This study evaluated the stability of models under different mask ratios of input and examines the correlation between the model's self-reported confidence level and objective accuracy.
We propose the following hypotheses:
\begin{enumerate}
    \item \textbf{Confidence-accuracy miscalibration}: The self-reported confidence level of LLMs deviates from their actual accuracy, often exhibiting overconfidence or overcaution, which may reduce their effectiveness in instructional guidance.
    \item \textbf{Perturbation sensitivity}: Model accuracy declines as the proportion of randomly masked content increases. This expected performance decline reflects differences among models in their ability to recover context and their tolerance for informational fragmentation.
\end{enumerate}

Understanding these characteristics will contribute to the safer and more efficient integration of LLMs into computer science education.

Following this introduction, Section~\ref{sec:methods} outlines the experimental setup and assessment criteria. Section~\ref{sec:results} reports the benchmark outcomes, which are then synthesized into broader educational implications in Section~\ref{sec:discussion}. Concluding the study, Section~\ref{sec:conclusion} offers a vision for future research and practical applications in CS pedagogy.

\section{Methodology}
\label{sec:methods}

\subsection{Large Language Models}
This study evaluated four prominent LLMs, selected based on three key criteria: architectural diversity, superior natural language capabilities, and accessibility via official APIs.

The selected models are as follows:
\begin{description}[leftmargin=*,labelindent=0em]
    \item[GPT-5 (OpenAI, 2025):] The flagship model from OpenAI, released on August 7, 2025. It demonstrates advances in reasoning, long-context processing, and multimodal capabilities \citep{openai2025gpt5}.
    \item[DeepSeek-R1 (DeepSeek AI, 2025):] An open-source reasoning-oriented model released on January 20, 2025, optimized for mathematical and programming tasks \citep{guo2025deepseek}.
    \item[Llama-3.3-70B-Instruct (Meta AI, 2024; hereafter Llama-3.3):] A 70-billion-parameter instruction-tuned model released on December 6, 2024, with improved multilingual support and reasoning proficiency \citep{meta2024llama33}.
    \item[Qwen-Plus (Alibaba Cloud, 2025):] A balanced variant from the Qwen3 series with iterative updates throughout 2025. It supports advanced step-by-step reasoning and demonstrates strong multilingual and coding performance \citep{qwen2025}.
\end{description}

To ensure reproducibility and isolate intrinsic capabilities, no fine-tuning or domain-specific adaptation was performed. All experiments utilized the models' official pre-trained and instruction-tuned configurations via their respective APIs.

\subsection{Dataset Construction and Assessment}
\label{sec:assessment}

We developed a multi-source benchmark dataset to rigorously evaluate leading LLMs in computer science education. The selection criteria for this question bank were primarily based on the official syllabi and competency standards of the six major certification exams. To accommodate the text-only models, all image-dependent questions were removed, resulting in a final question bank containing 1,068 text-only questions. These questions were primarily multiple-choice, with a few requiring multiple selections. They covered major areas such as computer networks (CCNA, Chinese Network Engineer), programming skills (OCJP, NCRE Java), and office application skills (ICDL, NCRE MS Office), ensuring high relevance and representativeness by covering a wide range of topics from basic to advanced.

Model responses, including final answers and reasoning processes, were collected from DeepSeek-R1, Qwen-Plus, Llama-3.3-70B-Instruct, and GPT-5. A strict scoring protocol was applied uniformly: both single-choice and multiple-choice responses were evaluated based on exact matches with official answer keys, with no partial credit awarded for incomplete or partially correct selections in multiple-choice questions.

\begin{figure}[!t]
    \centering
\includegraphics[width=\textwidth]{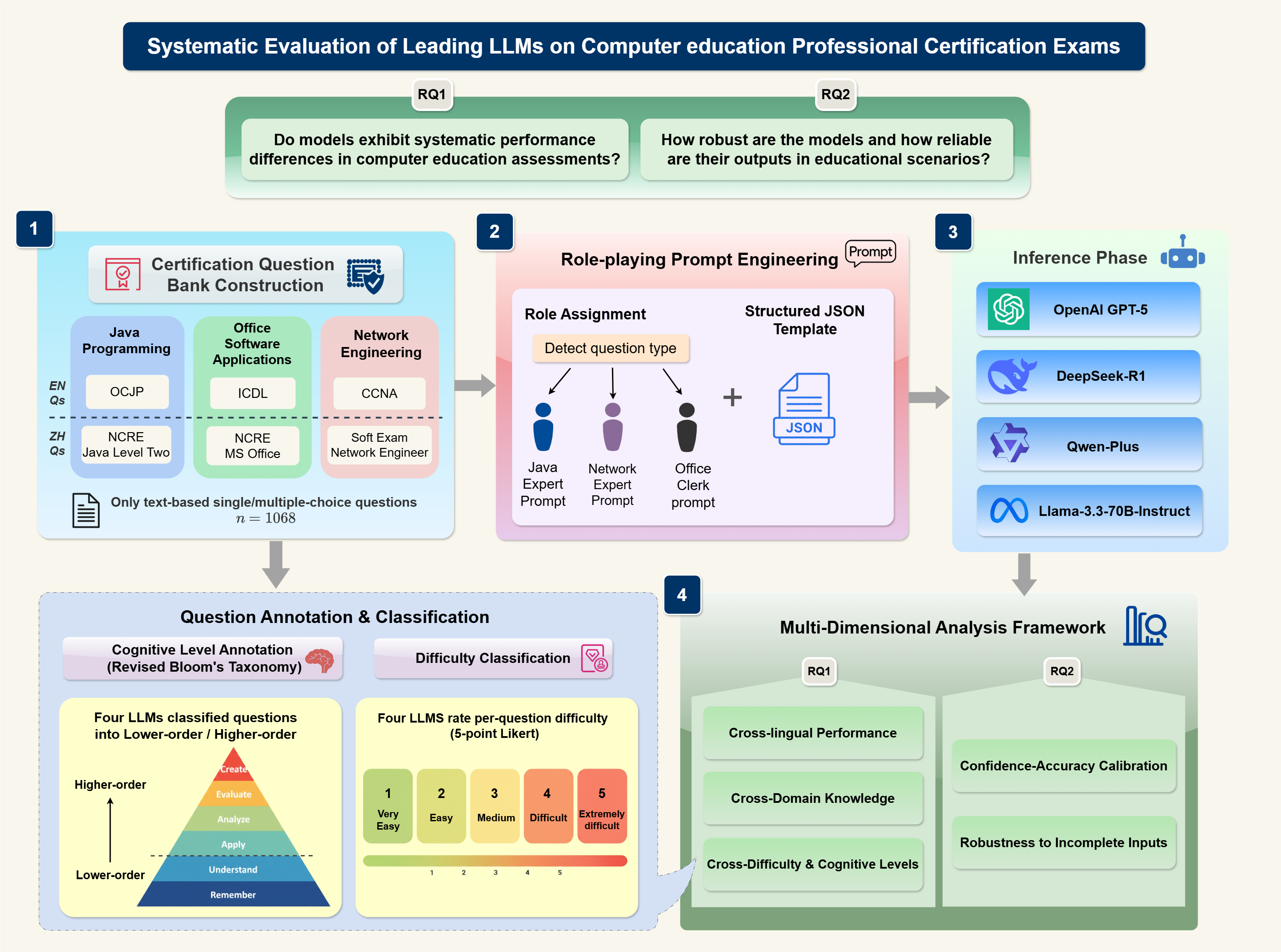}  
    \caption{The Overall Research Framework}
    \label{fig:framework}
\end{figure}

This multi-source design enabled relatively fair comparisons across certification systems while preserving the standardized assessment characteristics of real-world educational contexts, providing an educationally meaningful foundation for cross-model evaluation. The overall research framework is illustrated in Figure~\ref{fig:framework}.

\subsection{Prompt Engineering}

\begin{figure}[!t]
    \centering
    \includegraphics[width=1\textwidth]{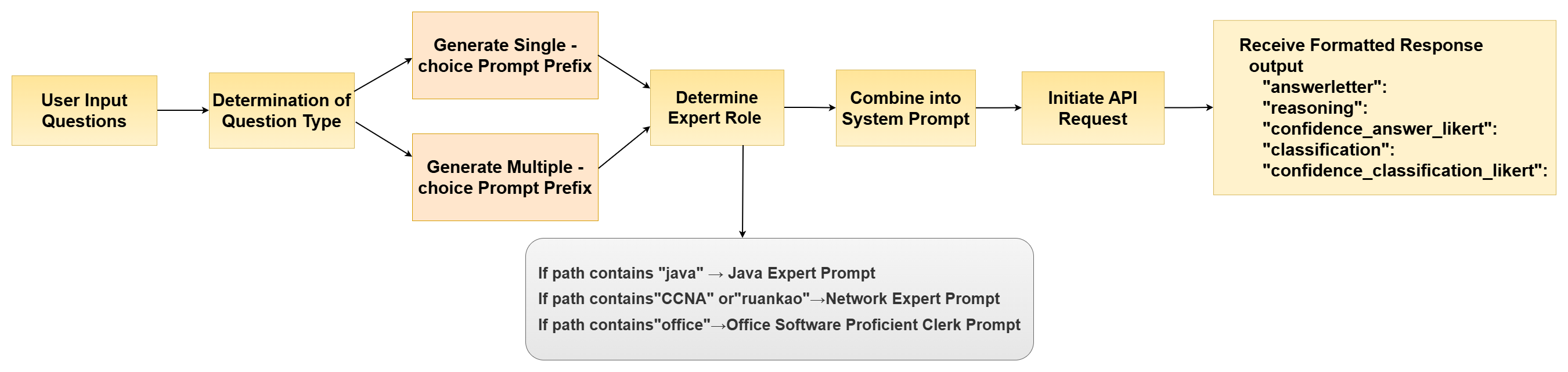}
    \caption{The Prompt Construction Process}
    \label{fig:process}
\end{figure}

Due to the significant impact of prompt engineering on the output of large language models (LLMs), we standardized the prompt structure and output format. Specifically, depending on the question type, we assigned corresponding role definitions to the model (e.g., Java Expert, Network Expert, or Office Software Proficient Clerk), requiring it to invoke relevant domain knowledge and provide accurate answers with detailed reasoning explanations. For each question, whether single-choice or multiple-choice, the question stem and all answer options were supplied to the model via its API. The detailed system prompts, including role definitions and output instructions, are provided in Appendix~\ref{app:system-prompts}. The overall prompt construction process is illustrated in Figure~\ref{fig:process}.

\subsection{Cross-lingual Evaluation}

We constructed a parallel Chinese-English corpus using DeepSeek-V3 to automatically translate the original question set~\citep{deepseekai2024deepseekv3technicalreport}. This aimed to investigate how language affects the effectiveness of the models and to examine potential cross-linguistic biases. The dataset consisted of three examination banks in Chinese (NCRE MS Office module, NCRE Java Level 2, and Chinese Network Engineer Examination) and three examination banks in English (ICDL, OCJP, and CCNA).

The model was positioned as a professional translator specializing in computer science and information technology by generating translations via the official DeepSeek-V3 API with a standardized system prompt (full prompt in Appendix~\ref{app:system-prompts}). This prompt explicitly instructed the model to maintain the original multiple-choice format, including question stems, options, and contextual descriptions, as well as to preserve strict semantic fidelity and technical accuracy, retain domain-specific terminology (e.g., networking protocols, Java syntax, and office software functions), and adhere to the original format. Furthermore, it directed the model to generate natural and idiomatic expressions in the target language, avoiding literal translations that might induce ambiguity, while preferring the preservation of the original purpose and complexity.

Translations were conducted in two distinct directions: from Chinese originals to English and from English originals to Chinese. The research team manually verified the translated items to guarantee translation quality and reduce errors from automated processing. Minor revisions were implemented solely to rectify evident semantic deviations or inconsistencies in critical technical terms. This method preserved the characteristics of a predominantly automated corpus while maintaining high fidelity to the source text.

The assessment of these translation tasks followed the rigorous scoring criteria, standardized question wording, and answer format established in the previous sections. By comparing the source and translation texts, quantitative research was conducted to explore differences in cross-language performance disparities and the reasons for performance degradation.

\subsection{Cognitive Level Classification}

To examine differences in model performance across tasks of varying cognitive complexity, we classified all questions into higher-order and lower-order categories based on Bloom's revised taxonomy~\citep{anderson2001taxonomy}. Higher-order questions primarily involve advanced cognitive processes such as applying, analyzing, evaluating, and creating, typically requiring multi-step reasoning, concept integration, or complex problem-solving. Lower-order questions focus on remembering and understanding, emphasizing direct retrieval of factual knowledge and basic comprehension.

The classification of these problems was conducted independently by four models (DeepSeek-R1, Qwen-Plus, Llama-3.3, and GPT-5) using standardized prompt templates (see Appendix~\ref{app:system-prompts} for complete prompts). The final category of each question was decided by a majority vote. If the results of at least three models were the same, the problem was marked as higher-order or lower-order. In order to ensure the reliability of the final classification, items resulting in a 2:2 split were excluded. Of the original 1,068 questions, 860 were retained after excluding the ties (exclusion rate $\approx$19.5\%).

This consensus-based approach provided a solid basis for analyzing the performance differences between higher-order and lower-order tasks.

In addition, in order to explore the subjective perception and internal consistency of the models on the difficulty of the task, a five-point Likert scale (1 = very easy, 5 = extremely difficult; full instructions are provided in Appendix~\ref{app:system-prompts}) was used to collect the difficulty scores from the four models. These scores were independent supplements to the classification of cognitive levels. They could test the potential bias and the reliability of the models' own difficulty assessments.

\subsection{Confidence Scoring}
\label{sec:confidence-scoring}
All four models were required to provide a confidence score for each answer on a five-point Likert scale. The scale was defined as follows: 
1 = no confidence (representing a pure guess), 
2 = low confidence (suggesting high uncertainty), 
3 = moderate confidence (reasonable but with reservations), 
4 = high confidence (indicating probable correctness with sound reasoning), 
and 5 = very high confidence (implying certainty in both the answer and the provided reasoning).

This structured confidence scoring provided the basis for subsequent analyses of the relationship between self-reported confidence and objective accuracy.

\subsection{Perturbation Robustness Evaluation}

To evaluate model robustness to incomplete input information, we conducted a random masking perturbation experiment designed to simulate noise and information loss commonly encountered in educational settings (e.g., irregular student expressions or transcription errors).

For each question stem, lightweight tokenization was applied: Chinese text was segmented into individual characters, while English text (sequences of letters, digits, and underscores) was split into words. Punctuation, spaces, and common placeholders were excluded from the maskable token set to preserve the core question structure. Let $T = \{t_1, t_2, \dots, t_N\}$ denote the resulting sequence of $N$ maskable tokens.

For a given masking ratio $r \in \{0.0, 0.2, 0.4, 0.6\}$, the number of tokens to mask was computed as
\begin{equation}
m = \lceil r \cdot N \rceil,
\end{equation}
where $\lceil \cdot \rceil$ is the ceiling function. A random subset $M \subset T$ of size $m$ was selected uniformly without replacement. Each selected token $t_i \in M$ was replaced with the fixed placeholder ``[MASK]'', producing the perturbed sequence
\begin{equation}
T' = \{ t'_1, t'_2, \dots, t'_N \},
\end{equation}
where $t'_i = \text{[MASK]}$ if $t_i \in M$ and $t'_i = t_i$ otherwise. Question options and overall format remained unchanged, ensuring perturbed inputs retained the structure of valid multiple-choice questions.

The four masking ratios correspond to no perturbation ($r=0.0$, baseline), mild ($r=0.2$), moderate ($r=0.4$), and severe ($r=0.6$) information loss, enabling systematic analysis of performance degradation under increasing perturbation intensity.

\subsection{Statistical Analysis}

Bootstrap resampling was used to estimate 95\% confidence intervals for model accuracy rates, as it is a non-parametric method well-suited to binary outcomes (correct/incorrect) and finite samples without normality assumptions.

For overall accuracy across all $N$ items, 1,000 bootstrap samples were drawn with replacement from the complete set of binary scores $\{x_i\}_{i=1}^N$ ($x_i \in \{0,1\}$). Accuracy for each replicate $b$ was computed as
\[
\hat{p}^{(b)} = \frac{1}{N} \sum_{i=1}^N x_i^{(b)}.
\]
The 95\% percentile confidence interval was then obtained as
\[
CI_{95\%} = [P_{2.5\%}, P_{97.5\%}],
\]
where $P_{2.5\%}$ and $P_{97.5\%}$ are the 2.5th and 97.5th percentiles of the 1,000 bootstrap accuracy estimates.

The same procedure was applied independently to each certification question set $j$ (with $n_j$ items) to produce set-specific intervals:
\[
\hat{p}_j^{(b)} = \frac{1}{n_j} \sum_{k=1}^{n_j} x_{jk}^{(b)}, \quad CI_{95\%}^{(j)} = [P_{j,2.5\%}, P_{j,97.5\%}].
\]

Inter-rater agreement among the four models' perceived difficulty ratings (ordinal scale 1--5) was assessed using Krippendorff's $\alpha$, computed only on items with valid ratings from all models to appropriately handle missing data.

All bootstrap procedures used $B = 1,000$ iterations and a nominal confidence level of 95\% ($\alpha = 0.05$). Analyses were implemented in Python (version 3.10 or higher) using NumPy for bootstrap resampling and percentile computation.

\section{Results}
\label{sec:results}
\subsection{Overall Performance on Six Certification Examinations}
This study evaluated the performance of large language models (LLMs), including DeepSeek-R1, Qwen-Plus, Llama-3.3, and GPT-5, on multiple professional certification examination datasets, revealing significant differences in their capabilities across exams, domains, and linguistic contexts.

Among the six professional certification examinations evaluated, GPT-5 achieved the highest accuracy in four tasks: CCNA (93.2\% [89.8\%, 95.8\%]), ICDL (94.5\% [89.9\%, 98.2\%]), Chinese Network Engineer Examination (90.3\% [86.4\%, 93.8\%]), and OCJP (77.8\% [70.8\%, 84.1\%]) (see Table~\ref{tab:accuracy}). It consistently outperformed the second-best model by 1.2 to 3.5 percentage points on the English-dominant international certification examinations (CCNA, ICDL, and OCJP, respectively).

\begin{table}[!ht]
\centering
\setlength{\tabcolsep}{4pt}
\footnotesize

\caption{Accuracy (\%) and 95\% Confidence Intervals of Four Large Language Models on Six Professional Certification Examinations}
\label{tab:accuracy}

\begin{tabular}{l *{4}{c}}
\toprule
Examination & GPT-5 & DeepSeek-R1 & Qwen-Plus & Llama-3.3 \\
\midrule
CN Net.Eng.
  & 90.3 [86.4, 93.8]
  & 89.5 [85.6, 93.0]
  & 89.5 [85.6, 93.0]
  & 82.1 [77.4, 86.8] \\
CCNA
  & 93.2 [89.8, 95.8]
  & 92.0 [88.6, 94.7]
  & 87.9 [83.7, 91.7]
  & 82.6 [78.0, 87.1] \\
NCRE MS Office
  & 90.4 [84.0, 95.7]
  & 92.6 [87.2, 97.9]
  & 92.6 [87.2, 96.8]
  & 84.0 [76.6, 91.5] \\
ICDL
  & 94.5 [89.9, 98.2]
  & 91.7 [86.2, 96.3]
  & 85.3 [78.0, 91.7]
  & 89.0 [83.5, 94.5] \\
NCRE Java
  & 93.5 [90.0, 96.5]
  & 94.0 [90.5, 97.0]
  & 96.5 [93.5, 99.0]
  & 85.5 [80.5, 90.0] \\
OCJP
  & 77.8 [70.8, 84.1]
  & 74.3 [67.4, 81.2]
  & 72.2 [64.6, 79.2]
  & 61.8 [54.8, 70.1] \\
\bottomrule
\end{tabular}

\smallskip
{\scriptsize \textit{Note}: Accuracy and confidence intervals are reported as percentages (rounded to one decimal place). Confidence intervals are approximate and derived via statistical resampling methods (e.g., bootstrap).}
\end{table}

Qwen-Plus demonstrated strong performance on Chinese-original domestic examinations, attaining the highest accuracy on NCRE Java (96.5\% [93.5\%, 99.0\%]), surpassing GPT-5 by 3.0 percentage points. On NCRE MS Office, Qwen-Plus tied with DeepSeek-R1 at 92.6\% [87.2\%, 96.8\%], outperforming GPT-5 by 2.2 percentage points.

DeepSeek-R1 exhibited strong cross-task and cross-lingual consistency, ranking in the top two across all six examinations. It attained at least 89.5\% accuracy in five examinations ($\geq$90.0\% in four: CCNA 92.0\% [88.6\%, 94.7\%], NCRE MS Office 92.6\% [87.2\%, 97.9\%], ICDL 91.7\% [86.2\%, 96.3\%], and NCRE Java 94.0\% [90.5\%, 97.0\%]).

Llama-3.3 trailed the leading model in five of the six examinations (ranking third on ICDL), with performance gaps relative to the best-performing model per task ranging from 5.5 to 16.0 percentage points. The largest gap occurred on OCJP (16.0 percentage points behind GPT-5 at 61.8\% [54.8\%, 70.1\%]).

However, a substantial performance decline was observed across all models on OCJP (average accuracy 71.5\%). Subsequent difficulty coefficient assessments indicated that this examination was more challenging than the others. Even the leading model on this task, GPT-5, achieved only 77.8\% [70.8\%, 84.1\%] (15.7 percentage points lower than its 93.5\% [90.0\%, 96.5\%] on NCRE Java). The decline was most pronounced for Qwen-Plus (-24.3 percentage points to 72.2\% [64.6\%, 79.2\%]), followed by Llama-3.3 (-23.7 percentage points to 61.8\% [54.8\%, 70.1\%]) and DeepSeek-R1 (-19.7 percentage points to 74.3\% [67.4\%, 81.2\%]). The consistent pattern observed across models indicates that OCJP possesses a higher inherent difficulty than NCRE Java, a disparity likely stemming from factors beyond linguistic differences. While both exams are topically comparable, NCRE Java focuses more heavily on a domestic curriculum. 

Overall, by late 2025, the four evaluated LLMs demonstrated high accuracy in professional certification tasks, with most models surpassing the 80\% threshold in the majority of exams. However, performance varied systematically by language environment and examination type: GPT-5 achieved the highest scores in English-based international certifications, Qwen-Plus excelled in Chinese domestic certifications, and DeepSeek-R1 proved to be the most stable across diverse domains, despite underperforming in English-oriented programming tasks.

\begin{figure}[!t]
    \centering
    \includegraphics[width=\textwidth]{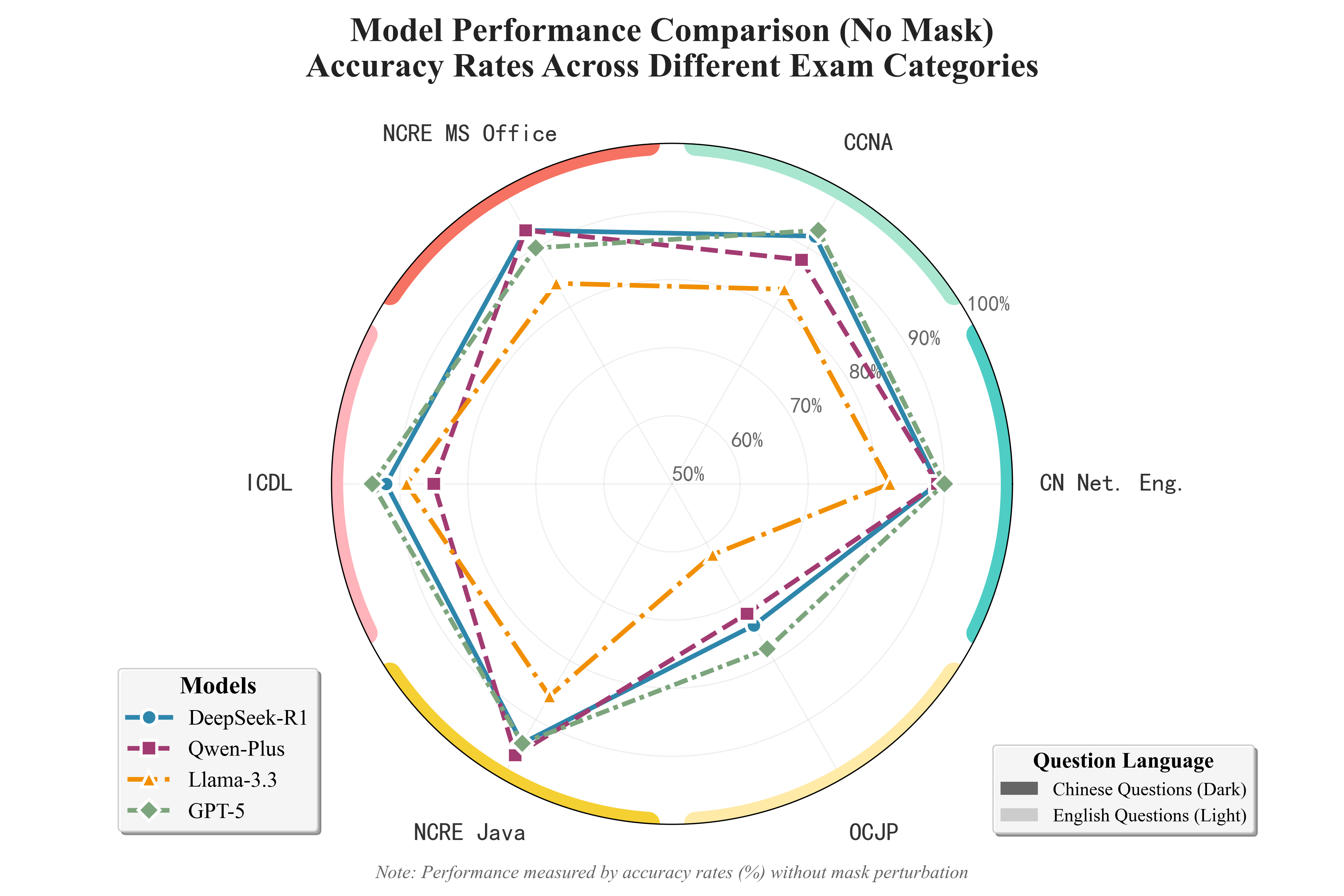}  
    \caption{Radar chart of model accuracy rates (no masking) across six professional certification examinations.}
    \label{fig:radar-no-mask}
\end{figure}

Figure~\ref{fig:radar-no-mask} provides a comprehensive visual summary of the models' accuracy rates across the six examinations without masking perturbation, highlighting the systematic differences in performance profiles across exam categories and linguistic contexts.

\subsection{RQ1: Systematic Differences in Computer Education Assessments Across Types and Attributes}
\subsubsection{Cross-lingual Performance and Directional Asymmetry}
\label{subsec:cross-lingual}

To examine the influence of language on model performance and probe potential cross-lingual biases, we constructed parallel Chinese-English corpora through professional bidirectional translation of the original question sets from the six certification examinations (three Chinese-original and three English-original). This yielded a cross-lingual evaluation dataset spanning computer networking, office applications, and programming skills.

\begin{figure}[!t]
    \centering
    
    \begin{subfigure}{0.48\textwidth}
        \centering
        \includegraphics[width=\textwidth]{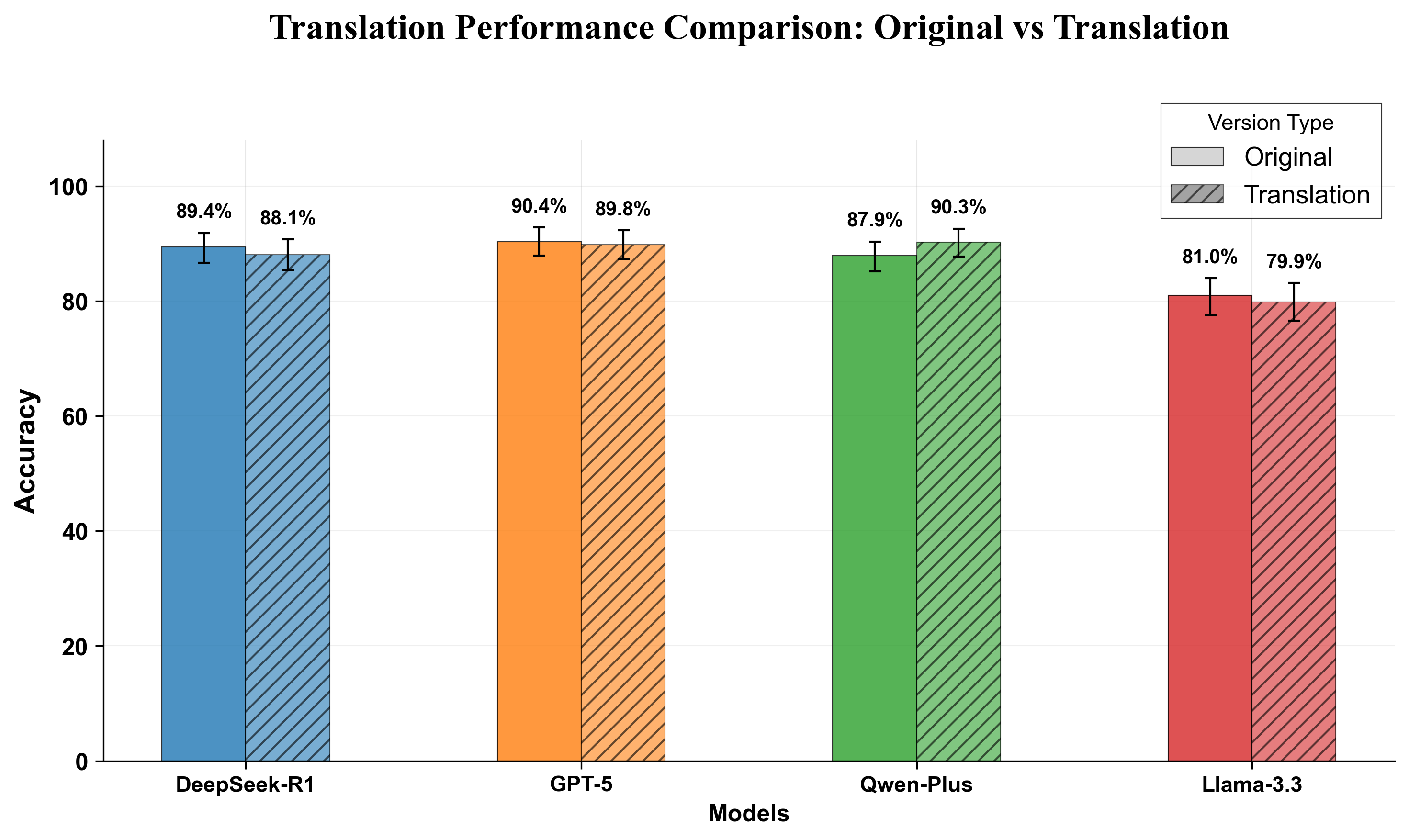}  
        \caption{Overall accuracy on original versus translated questions across models}
        \label{fig:translation-overall-a}
    \end{subfigure}
    \hfill  
    \begin{subfigure}{0.48\textwidth}
        \centering
        \includegraphics[width=\textwidth]{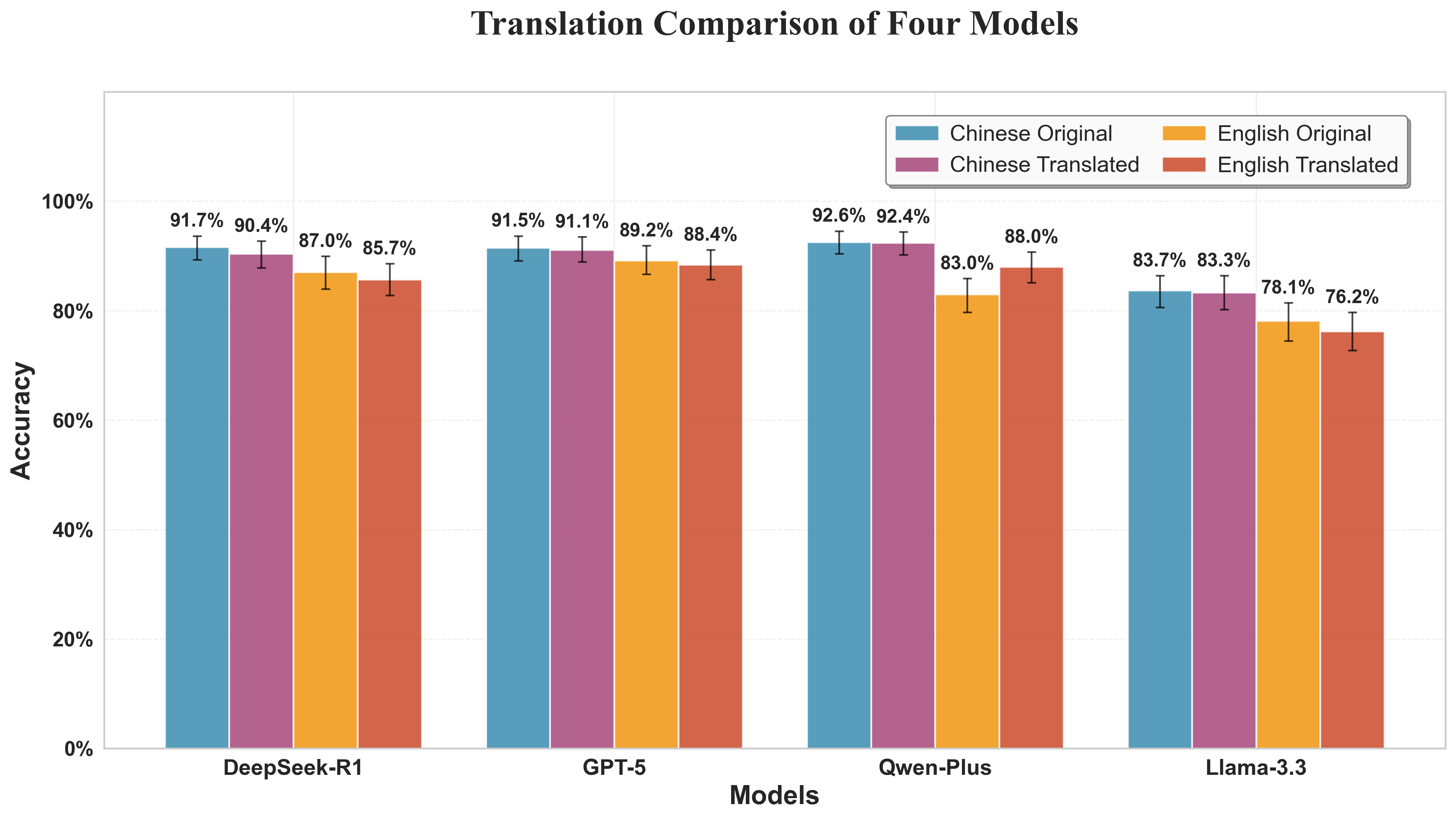}  
        \caption{Accuracy decomposed by question origin and translation direction (Zh$\to$En and En$\to$Zh)}
        \label{fig:translation-direction-b}
    \end{subfigure}
    
    \caption{Comparison of model performance on original versus translated questions. (a) Overall accuracy; (b) Decomposition by translation direction.}
    \label{fig:translation-comparison}
\end{figure}

Figure~\ref{fig:translation-comparison} presents a detailed comparison of model accuracy on original versus translated versions. Subfigure (a) shows overall performance across models, while subfigure (b) decomposes the results by question origin and translation direction.

As shown in Figure~\ref{fig:translation-comparison}(a), average accuracy on translated versions revealed distinct patterns of cross-lingual robustness. GPT-5 exhibited high stability, with a minimal decline of 0.6 percentage points (90.4\% [88.5\%, 92.1\%] to 89.8\% [87.4\%, 92.1\%]). DeepSeek-R1 showed a slight decline of 1.3 percentage points (89.4\% [87.6\%, 91.0\%] to 88.1\% [85.4\%, 90.7\%]). Llama-3.3 displayed a decline of 1.1 percentage points (81.0\% [78.8\%, 83.3\%] to 79.9\% [76.6\%, 83.2\%]). Qwen-Plus was the only model to show an overall improvement of approximately 2.3 percentage points (87.9\% [86.0\%, 90.0\%] to 90.3\% [87.7\%, 92.6\%]). This net gain stems from a pronounced Chinese language preference, as substantial gains in one translation direction outweighed minimal changes in the other (detailed below).

Further analysis by translation direction (Figure~\ref{fig:translation-comparison}(b)) reveals clear asymmetry between Chinese-to-English (Zh$\to$En) and English-to-Chinese (En$\to$Zh) directions. In the Zh$\to$En direction, all models demonstrated high stability with only minor declines: Qwen-Plus declined by 0.2 percentage points (92.6\% [90.4\%, 94.6\%] to 92.4\% [90.2\%, 94.4\%]), GPT-5 by 0.4 percentage points (91.5\% [89.1\%, 93.7\%] to 91.1\% [88.9\%, 93.5\%]), DeepSeek-R1 by 1.3 percentage points (91.7\% [89.3\%, 93.7\%] to 90.4\% [87.8\%, 92.7\%]), and Llama-3.3 by 0.4 percentage points (83.7\% [80.6\%, 86.4\%] to 83.3\% [80.2\%, 86.4\%]).

In contrast, the En$\to$Zh direction exhibited greater variation: Qwen-Plus improved substantially by 5.0 percentage points (83.0\% [79.7\%, 85.9\%] to 88.0\% [85.1\%, 90.7\%]), while DeepSeek-R1 declined by 1.4 percentage points (87.0\% [84.0\%, 89.9\%] to 85.7\% [82.8\%, 88.6\%]), GPT-5 by 0.8 percentage points (89.2\% [86.7\%, 91.9\%] to 88.4\% [85.7\%, 91.1\%]), and Llama-3.3 by 1.9 percentage points (78.1\% [74.5\%, 81.4\%] to 76.2\% [72.7\%, 79.7\%]).

Cross-directional comparison highlights divergent generalization capabilities. GPT-5 demonstrated the most balanced cross-lingual robustness, with small fluctuations in both directions (average absolute change $\approx$ 0.6 percentage points), exemplifying strong language-agnostic reasoning. Qwen-Plus showed a pronounced Chinese language preference, achieving near-peak performance ($\approx$ 92\%) on any Chinese-presented questions (original or translated from English), while starting from a lower baseline on native English questions; its substantial gain in the En$\to$Zh direction accounted for the overall improvement. DeepSeek-R1 exhibited consistent mild declines across directions, indicating balanced but moderate translation sensitivity. Llama-3.3, the weakest performer overall, experienced a larger decline in the En$\to$Zh direction. This indicates that when translated into its less-dominant language, it is particularly susceptible to semantic changes.

Overall, translation induced slight interference to most models, but had constructive benefits for some models (especially Qwen-Plus in the $\text{En} \to \text{Zh}$ direction) in a specific direction, which highlighted the difference in cross-language transfer efficacy between large language models.

\subsubsection{Cross-Domain and Subtopic Performance Differences}

\begin{figure}[!t]
    \centering
    \includegraphics[width=\textwidth]{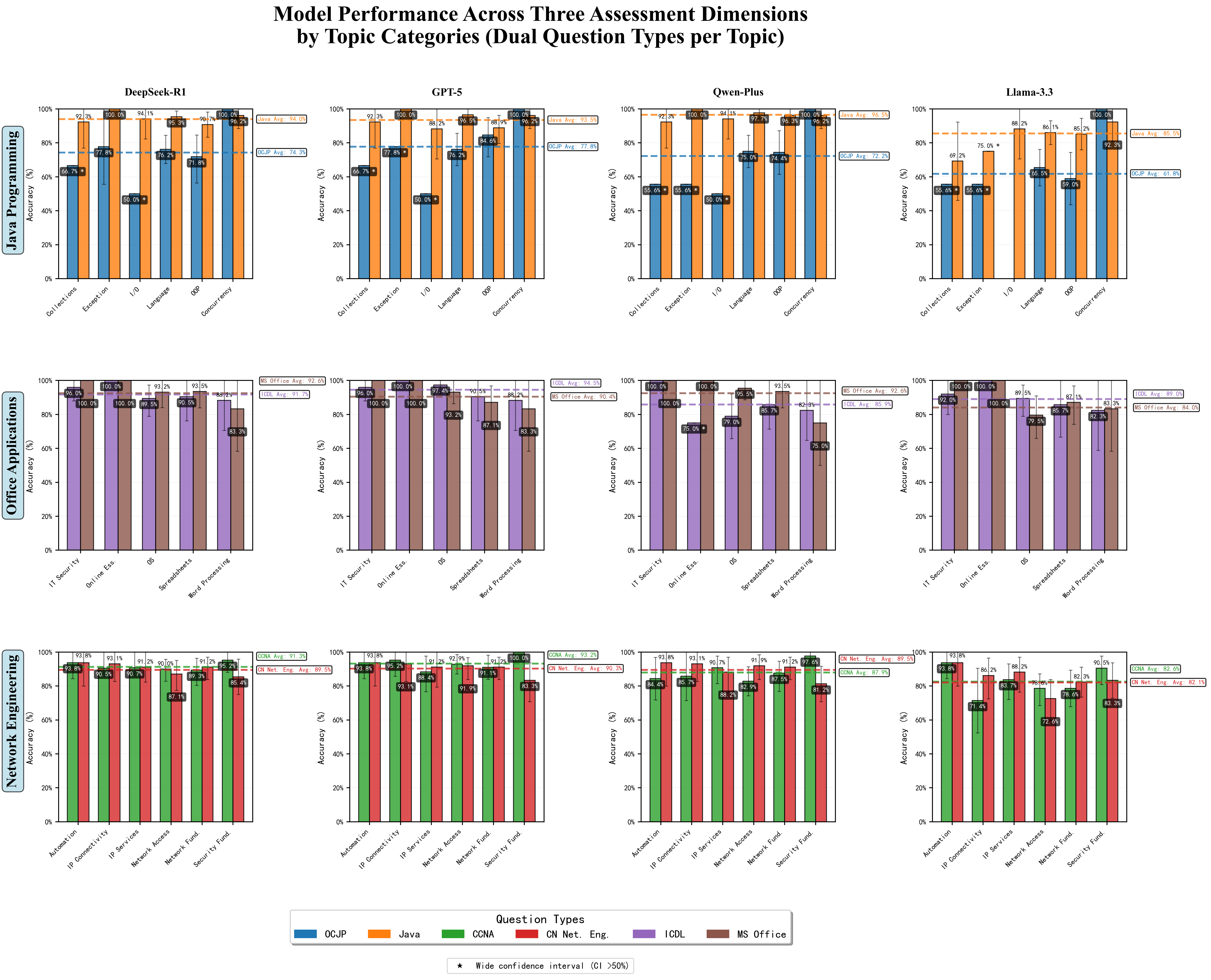}
    \caption{Model accuracy across domains and key subtopics with 95\% confidence intervals, showing divergence in complex reasoning-intensive tasks (e.g., OCJP OOP, MS Office OS, CCNA IP Connectivity).}
    \label{fig:cross-domain-subtopic}
\end{figure}

To evaluate the performance differences of four models across various domains and topics, we divided the six question sets into three main domains: Programming (Java), Office Applications, and Network Engineering, and further subdivided them into subtopics for a detailed cross-task performance analysis. Figure~\ref{fig:cross-domain-subtopic} provides a comprehensive visual summary of model accuracy across major domains and key subtopics, highlighting the observed divergences with 95\% confidence intervals.

In the Programming (Java) domain, model performance exhibited significant differences across various subtopics. In the Object-Oriented Programming (OOP) subtopic, the four models exhibited significant gaps: On the National Computer Rank Examination (NCRE) Java OOP question set, Qwen-Plus achieved a high accuracy of 96.3\% [90.7\%, 100.0\%], while Llama-3.3 scored 85.2\% [75.9\%, 94.4\%]. However, on the more challenging Oracle Certified Java Programmer (OCJP) OOP question set, GPT-5 reached 84.6\% [71.8\%, 94.9\%], whereas Llama-3.3 achieved 59.0\% [43.6\%, 74.4\%]. Llama-3.3's primary errors were concentrated in questions requiring deep conceptual understanding and multi-step reasoning, such as complex inheritance, polymorphism implementation, encapsulation design, and exception handling. This suggests relative weaknesses in Llama-3.3's reasoning about advanced OOP mechanisms.

In the Office Applications domain, all models performed at high levels on basic IT security knowledge. However, notable divergence emerged in advanced subtopics such as Operating Systems, Spreadsheets, and Word Processing. These tasks require models to accurately interpret user intent and map it to specific multi-step operational procedures, imposing greater demands on depth of functional knowledge and precision in logical reasoning. In the "Operating System" subtopic, the models exhibited differences across two independent test sets: the ICDL certification question bank and the MS Office certification question bank. GPT-5 achieved high accuracy of 97.4\% [92.1\%, 100.0\%] on the ICDL question bank and 93.2\% [86.4\%, 100.0\%] on the MS Office question bank, demonstrating comprehensive knowledge coverage and stable cross-bank performance. DeepSeek-R1 performed consistently, scoring 89.5\% [79.0\%, 97.4\%] and 93.2\% [84.1\%, 100.0\%] on the two banks, respectively. Notably, Qwen-Plus exhibited a pronounced difference: it achieved 95.5\% [88.6\%, 100.0\%] accuracy on the OS topic in the MS Office question bank but 79.0\% [65.8\%, 92.1\%] on the ICDL question bank, indicating limited coverage of internationally standardized certification systems. In contrast, Llama-3.3 showed the reverse pattern, achieving 89.5\% [79.0\%, 97.4\%] on the ICDL OS topic and 79.6\% [65.9\%, 90.9\%] on the MS Office question bank, suggesting greater alignment with general international concepts and relatively less depth in specific software ecosystem operations. These patterns suggest that the training data and optimization priorities of different models exhibited domain-specific biases.

Similar patterns were observed in the "Spreadsheets" and "Word Processing" subtopics, where GPT-5 and DeepSeek-R1 generally exhibited more stable performance in handling such complex, professional office tasks compared to the other models.

In the Network Engineering domain, most models performed at high levels across topics, while Llama-3.3 recorded the lowest accuracy in multiple key subtopics. Its relative weaknesses were evident in tasks requiring complex reasoning. In the core subtopic of IP Connectivity (CCNA questions), Llama-3.3 achieved 71.4\% [52.4\%, 90.5\%]. In comparison, the other three models performed higher: GPT-5 achieved 95.2\% [85.7\%, 100.0\%] accuracy, DeepSeek-R1 90.5\% [85.7\%, 100.0\%], and Qwen-Plus 85.7\% [71.4\%, 100.0\%]. IP Connectivity problems typically require multi-step logical reasoning to analyze routing paths in complex network topologies or diagnose connectivity faults; Network Access involves integrated application of knowledge in technology selection, protocol configuration, and security policies. Llama-3.3's consistently lower performance in these complex subtopics may stem from limited optimization and specialized training for vertical domains like network engineering. Its training corpus appears to cover basic network concepts sufficiently, but less so in knowledge representation and reasoning mechanisms for higher-order engineering problems requiring translation of theoretical principles into specific configurations and chain-like reasoning in complex scenarios. This observation aligns with Llama-3.3's relatively lower performance in similarly multi-step reasoning-intensive tasks, such as Programming (OCJP OOP) and advanced Office Applications functions. Collectively, these results, as visualized in Figure~\ref{fig:cross-domain-subtopic}, underscore its relative constraints in handling complex professional tasks compared to the other models.

\subsubsection{Performance Differences by Cognitive Question Level}
\label{subsec:cognitive-level}

To explore the relationship between model performance and the intrinsic cognitive demands of the questions, we implemented a majority-voting process among four LLMs based on the revised Bloom’s Taxonomy \citep{anderson2001taxonomy}. The 860 remaining questions were categorized into two groups: higher-order tasks ($n=36$) and lower-order tasks ($n=824$). For detailed procedures, please refer to Section~\ref{sec:methods}.

\begin{figure}[!t]
    \centering
    \includegraphics[width=\textwidth]{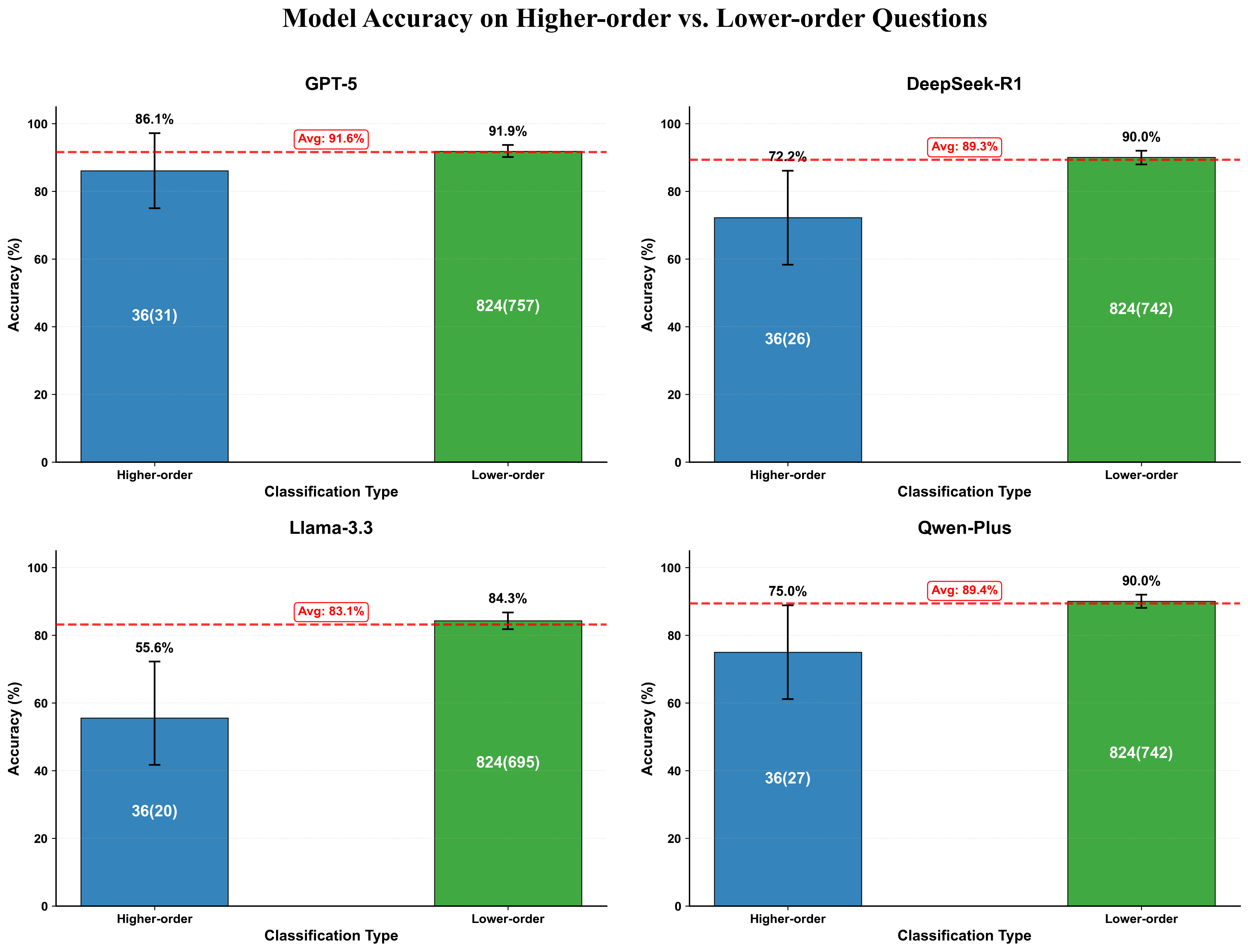} 
    \caption{Accuracy on higher-order (blue) and lower-order (green) questions for each model, based on Bloom's revised taxonomy. Dashed red lines show model-specific averages. Error bars indicate 95\% confidence intervals.}
    \label{fig:cognitive-gap}
\end{figure}

Figure~\ref{fig:cognitive-gap} illustrates that all models exhibit significant performance gaps between higher-order and lower-order questions, confirming our Hypothesis 3, that cognitive complexity has a significant impact on model performance.

GPT-5 achieved an accuracy of 86.1\% [75.0\%, 97.2\%] on higher-order tasks and 91.9\% [90.2\%, 93.7\%] on lower-order ones, with an overall average of 91.6\%. These figures underscore the model's exceptional cognitive resilience and stable reasoning, particularly its ability to maintain high performance under increased cognitive load. DeepSeek-R1 recorded 72.2\% [58.3\%, 86.1\%] on higher-order and 90.0\% [88.0\%, 92.0\%] on lower-order questions (average 89.3\%). This reflected balanced performance with moderate limitations in complex reasoning. Qwen-Plus attained 75.0\% [61.1\%, 88.9\%] on higher-order and 90.0\% [88.0\%, 92.0\%] on lower-order questions (average 89.4\%). It exhibited exceptional knowledge retrieval on lower-order tasks and slightly outperformed DeepSeek-R1. However, the noticeable decline on higher-order questions suggested potential for improvement in advanced cognitive integration.Regarding Llama-3.3, while it achieved 84.3\% on lower-order tasks, its accuracy plummeted to 55.6\% on higher-order questions. This 28.7-point drop—the largest observed in this study—reveals fundamental constraints in the model's capacity for complex reasoning and abstract information processing.

This consistent divergence highlighted the sensitivity of current model architectures and training strategies to cognitive hierarchies. Higher-order questions typically required inference of implicit assumptions, integration of cross-domain knowledge, or scenario-based simulation. These demands placed greater pressure on contextual modeling and long-chain reasoning capabilities. GPT-5's relative advantage likely stemmed from its advanced dynamic reasoning mechanisms and multimodal pre-training. In contrast, Llama-3.3's pronounced lag reflected generalization deficiencies in higher-order tasks within specialized professional domains. These patterns aligned with earlier cross-task observations, such as Llama-3.3's low performance on complex OOP inheritance and network IP connectivity problems. They reinforced the need for targeted architectural or training enhancements to better support higher-order cognitive demands in educational AI systems.

As a supplementary exploratory analysis, perceived difficulty ratings were collected from the four models on a five-point Likert scale (1 = Very Easy, 5 = Extremely Difficult; prompt template in Appendix A). Inter-model agreement on absolute ratings was low (overall Krippendorff's $\alpha = 0.13$; per-category range: -0.16 to 0.17), which reflected limited convergence in the models' difficulty judgments (see Table~\ref{tab:appendix-difficulty-ratings} in Appendix~\ref{app:difficulty-ratings} for details).

DeepSeek-R1 consistently assigned low difficulty scores (mean = 1.34 across categories). Qwen-Plus tended toward higher ratings (mean = 2.17), while Llama-3.3 fell in an intermediate range (mean = 2.12) and GPT-5 showed relatively low ratings overall (mean = 1.40). Despite the low absolute agreement, all models converged on the same relative ordering of question banks: OCJP was consistently ranked as the most challenging, followed by networking-related categories, while office application categories were perceived as the easiest. This shared ordinal perception suggested that LLMs could capture consistent relative task complexity patterns even when absolute difficulty judgments diverged. Given the low inter-model reliability, perceived difficulty scores were not used as a primary explanatory variable. Performance differences were more robustly captured by observed accuracy rates and the multi-model voting-based cognitive classification.

\subsection{RQ2: Robustness and Reliability Assessment}
\label{subsec:rq2}

\subsubsection{Relationship Between Confidence Levels and Accuracy}
\label{subsec:confidence-accuracy}

\begin{figure}[!t]
    \centering
    
    \begin{subfigure}{\textwidth}
        \centering
        \includegraphics[width=0.85\textwidth]{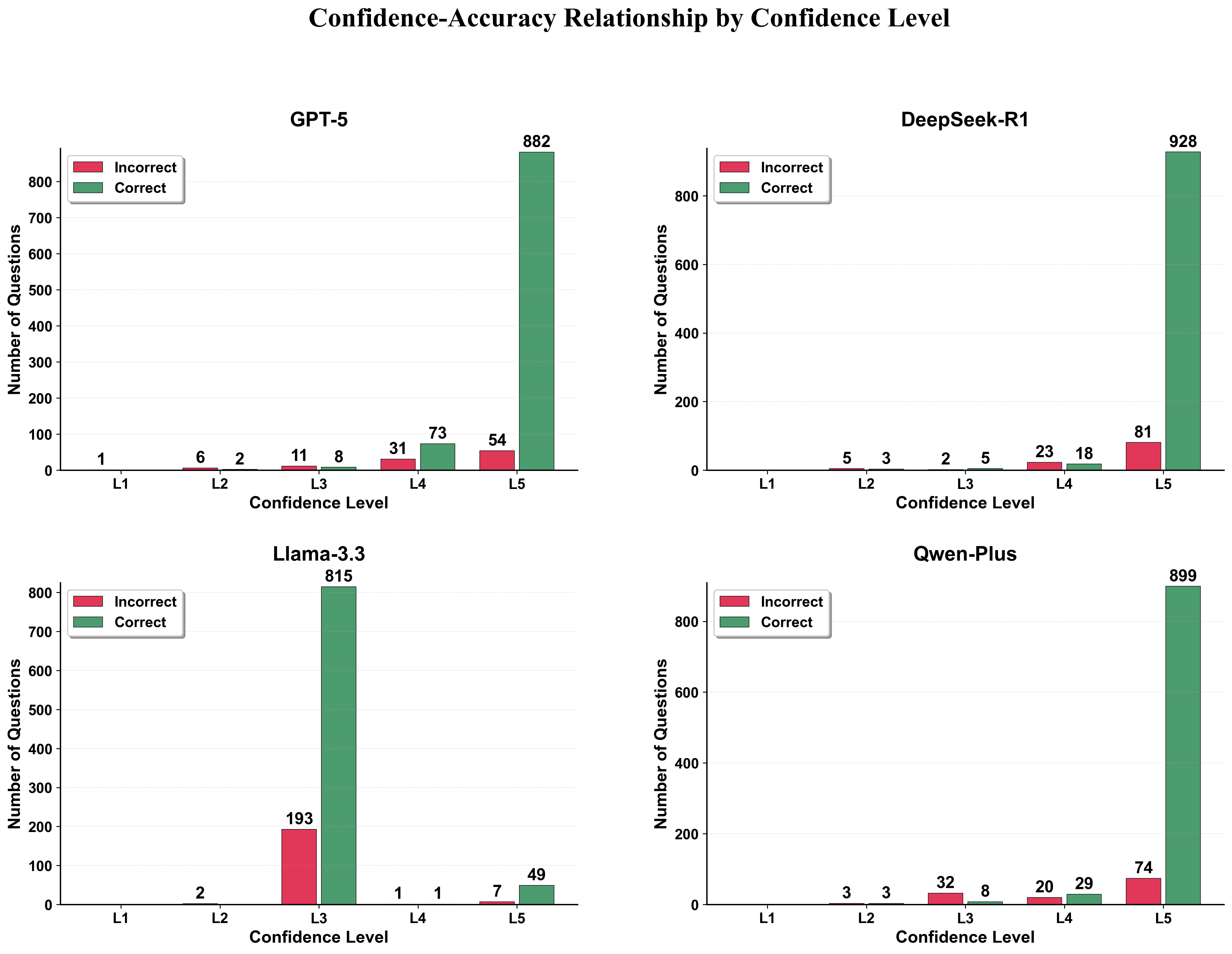}  
        \caption{Distribution of correct (green) and incorrect (red) responses across confidence levels (L1--L5) for each model}
        \label{fig:confidence-accuracy-a}
    \end{subfigure}
    
    \vspace{0.8em}  
    
    \begin{subfigure}{\textwidth}
        \centering
        \includegraphics[width=0.85\textwidth]{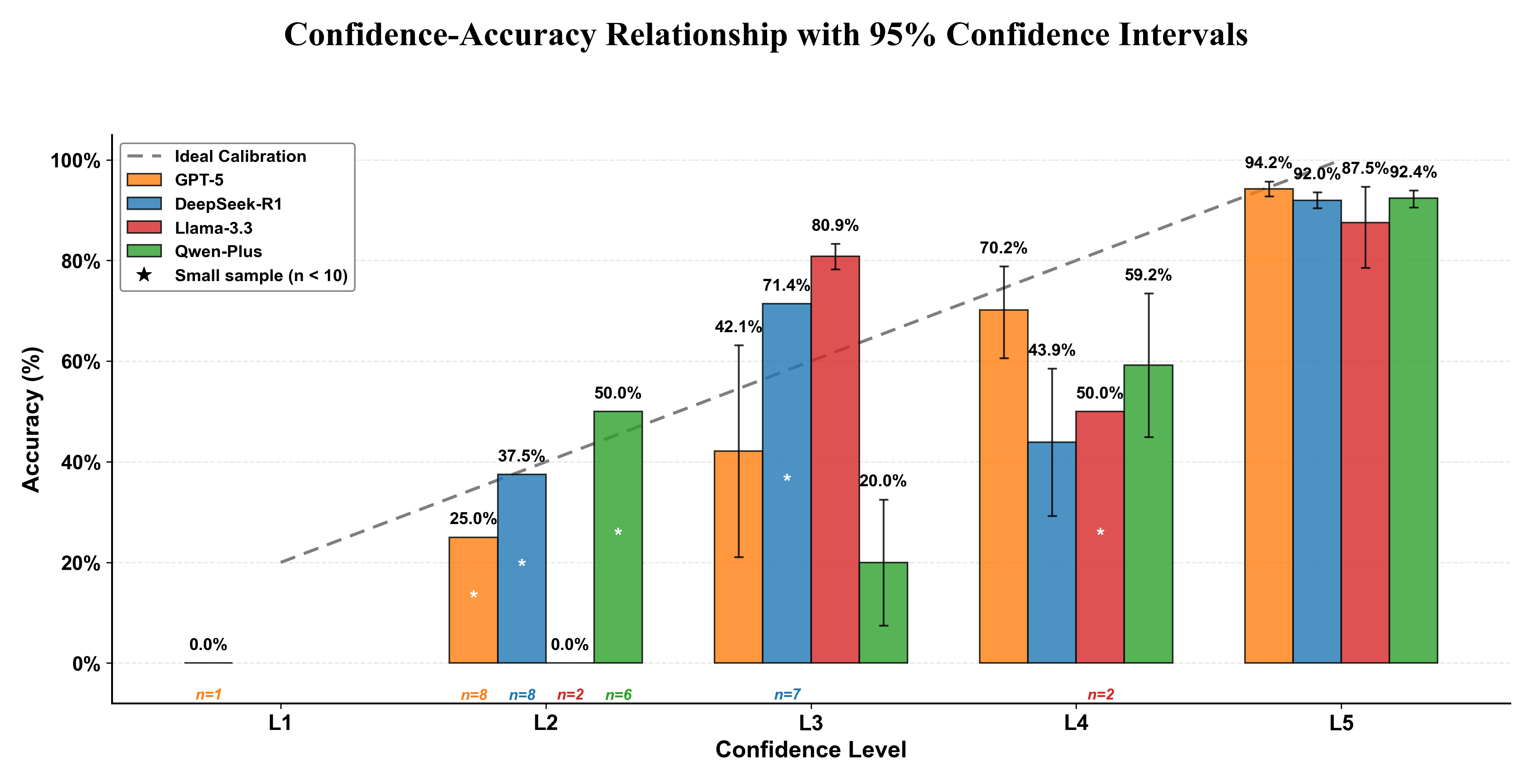} 
        \caption{Accuracy at each confidence level with 95\% confidence intervals. The dashed gray line represents ideal calibration. Asterisks indicate small sample sizes (n < 10)}
        \label{fig:confidence-accuracy-b}
    \end{subfigure}
    
    \caption{Confidence-accuracy relationship by confidence level. (a) Distribution of correct and incorrect responses. (b) Accuracy with 95\% confidence intervals.}
    \label{fig:confidence-accuracy}
\end{figure}

To evaluate the reliability of model outputs, we examined the relationship between self-reported confidence levels (L1 to L5 on a five-point Likert scale) and actual answer correctness across all questions. Figure~\ref{fig:confidence-accuracy} presents this relationship in two complementary views.

Figure~\ref{fig:confidence-accuracy}(a) shows the distribution of correct and incorrect responses at each confidence level for the four models. All models showed a positive correlation between confidence and accuracy, and as the level of confidence increased, the proportion of correct answers also increased. Responses at L5 (highest confidence) accounted for the largest number of correct answers for GPT-5 (882 out of 936), DeepSeek-R1 (928 out of 1009), and Qwen-Plus (899 out of 973). Lower confidence levels (L1 to L3) contained relatively few responses and a higher proportion of errors for these models.

In contrast, Llama-3.3 assigned the majority of responses to L3 (1,008 responses, 815 correct and 193 incorrect) and rarely assigned L5 (56 responses, 49 correct and 7 incorrect). This distribution was consistent with its lower overall performance and elevated baseline error rate. While Llama-3.3 delivered impressive accuracy even at L5, its tendency to avoid high-certainty declarations created a practical bottleneck. This rarity of 'confident' signals restricted the model's utility in high-stakes professional workflows where a definitive response is mandatory.

Figure~\ref{fig:confidence-accuracy}(b) quantifies the confidence-accuracy relationship, plotting accuracy at each confidence level with 95\% confidence intervals. All models showed increasing accuracy with higher confidence, approaching or exceeding 90\% at L5. GPT-5 demonstrated strong calibration, with an L5 accuracy of 94.2\% [92.7\%, 95.7\%]. DeepSeek-R1 and Qwen-Plus followed closely, with L5 accuracies of 92.0\% [90.4\%, 93.6\%] and 92.4\% [90.5\%, 93.9\%], respectively. Llama-3.3’s accuracy remained lower across the spectrum; while it achieved its highest accuracy of 87.5\% at L5, its performance at the most frequently assigned level (L3) stood at 80.9\% [78.3\%, 83.3\%]. Points based on small sample sizes ($n < 10$) are marked with asterisks to indicate higher uncertainty. Detailed confidence-accuracy intervals by confidence level are provided in Table~\ref{tab:appendix-confidence-intervals} in Appendix~\ref{app:confidence-intervals}.

Overall, the confidence-accuracy relationship observed for GPT-5, DeepSeek-R1, and Qwen-Plus indicated well-calibrated mechanisms, enabling educators to prioritize or filter responses based on reported confidence. Llama-3.3's conservative distribution and higher variability at lower levels constrained its reliability, particularly in automated tutoring or assessment contexts where clear, high-confidence thresholds are required for safe deployment.

\subsubsection{Robustness Under Information Missing}
\label{subsec:perturbation-sensitivity}

\begin{figure}[!t]
    \centering
    \begin{subfigure}{\textwidth}
        \centering
        \includegraphics[width=0.85\textwidth]{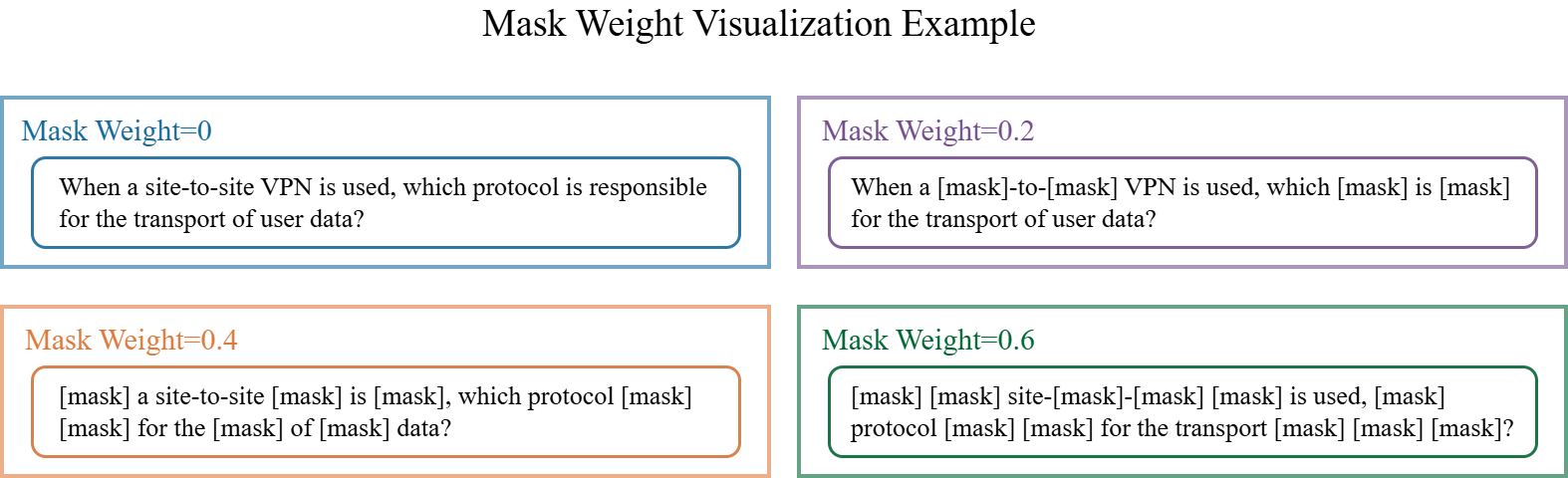}  
        \caption{Example of uniform random word-level masking applied to a sample question stem at different mask weights}
        \label{fig:mask-illustration-a}
    \end{subfigure}
    
    \vspace{0.8em}  
    
    \begin{subfigure}{\textwidth}
        \centering
        \includegraphics[width=0.85\textwidth]{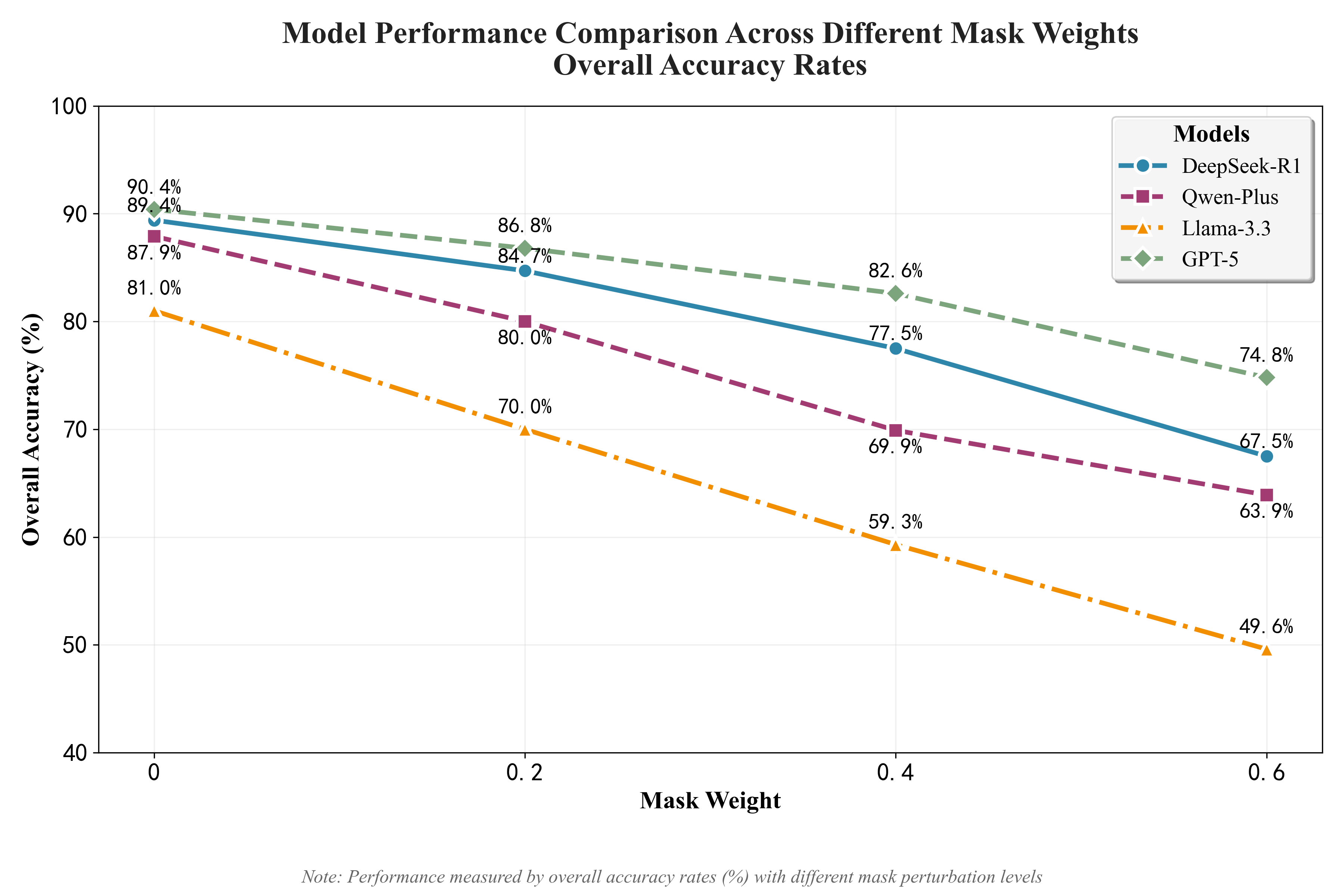}
        \caption{Overall accuracy of the four models as a function of mask weight}
        \label{fig:mask-performance-b}
    \end{subfigure}
    
    \caption{(a) Illustration of masking perturbation on a sample question stem. (b) Overall accuracy trends across mask weights. Accuracy in (b) is reported as the percentage of correctly answered questions under uniform random word-level masking.}
    \label{fig:mask-illustration}
\end{figure}

To simulate information loss, we subjected word-level tokens within the question stems to uniform random masking, testing across four intensity levels: 0.0, 0.2, 0.4, and 0.6. Figure~\ref{fig:mask-illustration}(a) provides an example of how the masking perturbation affects a sample question stem at different mask weights.

Accuracy decreased progressively with increasing mask weights across all models. GPT-5 showed the least pronounced decline, with accuracy decreasing from 90.4\% [88.6\%, 92.1\%] (no masking) to 74.8\% [72.3\%, 77.4\%] at 60\% masking (a 15.6 percentage point drop). DeepSeek-R1 and Qwen-Plus exhibited intermediate declines, from 89.4\% [87.5\%, 91.2\%] to 67.5\% [64.7\%, 70.5\%] (21.9 percentage point drop) and from 87.9\% [86.0\%, 89.9\%] to 63.9\% [61.0\%, 66.7\%] (24.0 percentage point drop), respectively. Llama-3.3 displayed the largest decline, from 81.0\% [78.8\%, 83.2\%] to 49.6\% [46.5\%, 52.7\%] at 60\% masking (31.4 percentage point drop), with a notable decrease of 11.0 percentage points already at 20\% masking.

Figure~\ref{fig:mask-illustration}(b) presents the overall accuracy trends across mask weights. GPT-5 consistently achieved the highest accuracy and the smallest reduction with increasing perturbation, whereas Llama-3.3 exhibited the largest reduction, particularly at higher masking levels.

These patterns indicate varying degrees of contextual recovery and inference capability under information loss. Specifically, GPT-5 exhibited superior resilience to masking perturbation, whereas Llama-3.3 demonstrated a more pronounced sensitivity to input completeness. These observed differences underscore the importance of examining the robustness of lower-performing models, particularly when evaluating AI systems for educational applications that may encounter noisy or fragmented inputs in real-world settings.

\section{Discussion}
\label{sec:discussion}
The study evaluates several late-2025 large language models using computer science certification exams as a test setting. Three domains are included—networking, programming, and office applications—allowing us to observe how performance varies under different conditions. In particular, we consider the role of language, domain-specific content, and cognitive demands. The main observations are reported first, followed by a discussion of possible reasons and areas that may require further study.

\subsection{Key Findings}
In response to RQ1, our findings demonstrate systematic performance gaps across language, domain, and cognitive dimensions, lending robust support to all three hypotheses.

In terms of language, we observed distinct model-specific dependencies: while GPT-5 dominated English-centric international certifications, Qwen-Plus led in Chinese domestic contexts, with DeepSeek-R1 maintaining the most balanced cross-lingual profile. We observed a sharp contrast in how translation affected performance. English-to-Chinese accuracy fluctuated widely, whereas accuracy in the Chinese-to-English direction showed a significantly narrower reduction. These trends highlight the importance of aligning the chosen model with the primary language of instruction when selecting a model in a multilingual environment.

Performance was notably uneven across subject domains, with distinct proficiency tiers emerging in programming, office applications, and network engineering. These gaps likely stem from differences in how each model was specialized or the breadth of its pre-training data—factors that make domain-matched deployment essential for balanced curriculum support. 

Cognitive demand had a substantial impact, with all models showing declines on higher-order tasks relative to lower-order ones. GPT-5 exhibited the smallest performance gap, while Llama-3.3 displayed the largest. Such a performance gap underscores the necessity of top-tier models when the goal is to foster higher-level cognitive skills—particularly analysis and creation—within the computer science curriculum.

In response to RQ2, the findings highlight significant gaps in how models handle calibration and robustness, which supports both hypotheses.

A positive correlation existed between confidence and accuracy for all models; higher confidence led to higher accuracy. GPT-5 exhibited the best calibration performance, followed by DeepSeek-R1 and Qwen-Plus. In contrast, Llama-3.3 employed a more conservative confidence distribution, with fewer high-confidence outputs and the majority of outputs concentrated in the medium-confidence range. These relationships suggest that confidence scoring can serve as a filtering mechanism to improve the reliability of artificial intelligence as an auxiliary educational tool.

Model sensitivity to input masking revealed a clear divide. GPT-5 was the most resilient when dealing with these masked prompts, whereas DeepSeek-R1 and Qwen-Plus maintained a moderate level of stability. In contrast, Llama-3.3 proved the most vulnerable to such disruptions. These differences suggest that while all models perform well under ideal conditions, their practical value in classrooms—where student input is often fragmented or messy—depends heavily on their individual robustness to incomplete data.

\subsection{Interpretation and Implications}

The findings for RQ1 extend general LLM benchmarks to structured educational contexts, revealing how training data distribution and architectural differences drive performance heterogeneity.

Observed language dependencies align with imbalances in multilingual pre-training corpora. Domestic models like Qwen-Plus appear to benefit from richer Chinese-language data, yielding enhanced performance when English content is translated into Chinese and compensating for lower baselines on native English questions. In contrast, international models like GPT-5 gain advantages from English-dominant corpora, supporting denser representations in English-centric tasks \citep{yang2024language,li2025impact}. The pronounced translation asymmetry---minimal interference in Chinese-to-English but greater variation in English-to-Chinese---suggests challenges in lower-resource language alignments \citep{kang2025language,liu2025conditions}. These patterns indicate opportunities for more balanced multilingual training to improve cross-lingual generalization in global educational applications.

Domain-specific unevenness is consistent with LLM specialization trends. Closed-source models benefit from diverse corpora for complex tasks (e.g., OOP and network connectivity), while open-source models like Llama-3.3 show limitations due to scaling and domain optimization constraints \citep{ling2025domain}. These variations suggest the potential value of domain-matched model deployment to ensure balanced coverage of curriculum topics.

The substantial declines on higher-order cognitive tasks confirm that current LLMs remain stronger in factual retrieval and application than in analysis, evaluation, and creation, aligning with gaps in higher Bloom's Taxonomy levels \citep{huber2025llms, jackson2025higher}. GPT-5's smaller gap suggests advantages in advanced reasoning mechanisms, whereas Llama-3.3's larger decline highlights persistent open-source challenges in sustained reasoning. This disparity indicates the benefit of prioritizing advanced models for higher-order learning objectives in computer science education.The supplementary analysis of perceived difficulty ratings revealed low inter-model agreement on absolute ratings (overall Krippendorff's $\alpha = 0.13$; per-category range: $-0.16$ to $0.17$), indicating substantial heterogeneity in how models perceive question difficulty. This limited convergence likely reflects differences in pre-training biases and internal representations of task complexity, consistent with prior observations of variability in LLM subjective judgments \citep{kadavath2022language,perez2022discovering,zheng2023judging}. Despite the low absolute agreement, all models converged on the same relative ordering of question banks, suggesting that LLMs can capture consistent ordinal task complexity patterns. These findings reinforce the prioritization of objective metrics (accuracy and multi-model voting) over self-reported difficulty and highlight challenges in using LLM subjective judgments for explanatory purposes in educational evaluations.

For RQ2, strong confidence-accuracy correlations were observed, with accuracy increasing at higher confidence levels, indicating reasonable calibration (where reported confidence approximates empirical accuracy). GPT-5 showed the tightest alignment, particularly at high confidence, followed by DeepSeek-R1 and Qwen-Plus. Llama-3.3 displayed a more conservative pattern, with under-confidence and fewer high-confidence outputs. This calibration supports practical applications, such as filtering high-confidence responses to enhance reliability in AI-assisted tutoring \citep{geng2024survey,kadavath2022language}.

Perturbation experiments revealed varying tolerance to input noise under masking. GPT-5's superior performance appears to stem from advanced contextual modeling and dynamic reasoning mechanisms, enabling effective inference from partial cues, particularly in programming and networking questions. DeepSeek-R1's marginal advantage over Qwen-Plus in some conditions may relate to its code and mathematical optimization. In contrast, Llama-3.3's greater vulnerability aligns with deficiencies in long-range dependency capture and noise robustness, where key term masking more readily disrupts coherence. These differences suggest advantages for resilient models like GPT-5 in real educational environments with unstable input quality (e.g., irregular student expressions or transcription errors) and indicate that robustness to random information loss merits consideration as a key evaluation metric for LLM deployment in practical settings \citep{alahmari2025large}.

Overall, while late-2025 LLMs demonstrate high proficiency in certification tasks, their heterogeneity in reasoning depth, cross-lingual transfer, input robustness, and calibration indicates the need for cautious integration in computer science education. Supervised deployment, combined with confidence filtering and domain matching, can help preserve student agency and promote equitable AI support.

\subsection{Limitations and Future Directions}

This study possesses certain limitations that offer guidance for future research endeavors.

First, the benchmark is limited to three domains: computer networks, office applications, and Java programming. Other important areas such as hardware fundamentals, databases, algorithms, cybersecurity, and AI ethics remain unexamined. Future research could extend the framework to cover these topics.

Second, this assessment focuses on multiple-choice questions from standardized certifications. While this facilitates objective and repeatable scoring, it neglects performance on open-ended tasks such as code debugging, system design, and technical writing. Incorporating generative and interactive assessment methods will be essential to more comprehensively measure the educational effectiveness of LLMs.

Third, the absence of human benchmarks limits this study to relative comparisons among the models. Benchmarking against actual test takers or domain experts would enable an absolute assessment of capabilities, thus more clearly defining the gap between current LLM proficiency and human expertise.

Fourth, as this study relies on static, single-round questions, it cannot capture the dynamic iterations inherent in real classroom interactions. Future research should prioritize longitudinal studies on real-time educational platforms to assess long-term consistency, responsiveness to student feedback, and robustness during sustained pedagogical dialogue.

Fifth, the cross-linguistic evaluation relied primarily on automated translation supplemented by limited human proofreading, which may have introduced subtle semantic shifts or terminological inconsistencies. Future work should employ professional human translation and rigorous expert cross-verification to construct more authoritative parallel corpora.

Beyond these methodological limitations, future research can address identified model shortcomings through targeted optimizations. These include developing specialized prompt engineering and Chain-of-Thought (CoT) strategies to enhance higher-order reasoning, and conducting domain-adaptive pre-training to bridge the performance gap between open-source and proprietary models in education-specific fields. Such advancements will bolster technological robustness and facilitate the equitable and secure integration of Large Language Models (LLMs) into global computer science education.

\section{Conclusion}
\label{sec:conclusion}
This investigation evaluated the efficacy of contemporary large language models within computer science professional certification frameworks. By benchmarking GPT-5, Qwen-Plus, DeepSeek-R1, and Llama-3.3, we identified critical performance disparities across diverse subject domains and linguistic contexts.

The empirical evidence reveals that while these models demonstrate proficiency in routine standardized tasks, their reliability fluctuates when confronted with domain-specific logic and higher-order reasoning. Notably, GPT-5 exhibited superior stability across various conditions, whereas Qwen-Plus showed a specialized advantage in Chinese-centric assessments. Conversely, Llama-3.3 displayed a pronounced sensitivity to task complexity and input fragmentation, highlighting a persistent "reasoning gap" in highly technical fields. These findings underscore that the integration of AI into educational evaluation must be context-dependent, prioritizing model robustness over general linguistic fluency. While certification tests provide a useful baseline, translating exam performance into effective classroom practice remains a distinct challenge. Further practical evidence is required to understand how these models truly affect student performance and learning in real-world settings.



\section*{Declaration of generative AI and AI-assisted technologies in the manuscript preparation process.}
This work originally studies the effects of LLMs in computer science education, with LLMs involved in the research design and methods. In addition, during the preparation of this manuscript, the author(s) used GPT 5.4 to improve language and correct typographical and grammatical errors. After using this tool/service, the author(s) reviewed and edited the content as needed and take(s) full responsibility for the content of the published article.

\bibliographystyle{elsarticle-harv}
\bibliography{references}

\newpage

\appendix

\section{System Prompts Used in the Study}
\label{app:system-prompts}

This appendix presents the complete system prompts employed in the experiments. These prompts were used for question answering with structured JSON output, cross-lingual translation, and question difficulty rating. All prompts enforce strict output formats for automated parsing and consistency.

\subsection{Expert Role Definitions}

The following expert role definitions are dynamically inserted into both the question answering prompt and the difficulty rating prompt based on keywords in the file path (e.g., ``Java'' or ``OCJP'' for Java Expert, ``CCNA'' or ``Network'' for Network Expert, ``ICDL'' or ``Office'' for Office Software Proficient Clerk). They provide domain-specific expertise to enhance reasoning accuracy.

\textbf{Java Expert:}
\begin{lstlisting}
You are a Java expert answering choice-based questions. Please analyze based on:
1. Code logic and output prediction
2. Syntax and performance issue identification
3. Core concept understanding (multithreading, collections, OOP)
4. Complex scenario analysis (inheritance, exceptions, generics)
Please carefully analyze the question and provide accurate answers.
\end{lstlisting}

\textbf{Network Expert:}
\begin{lstlisting}
You are a network expert answering choice-based questions. Please focus on:
1. Network protocols and standards
2. Routing and switching technologies
3. Network security and troubleshooting
4. Network design and optimization
Please provide accurate answers based on your expertise.
\end{lstlisting}

\textbf{Office Software Proficient Clerk:}
\begin{lstlisting}
You are an office worker who is proficient in office software applications and possesses extensive knowledge in this field.
You have many years of practical experience and are well-versed in various office software operations.
Now you need to answer the following questions based on your existing knowledge and expertise.
Please choose the most standardized and efficient operation plan.
\end{lstlisting}

\subsection{Question Answering and Structured Response Prompt}

The following prompt was used for answering single-choice and multiple-choice questions, requiring structured JSON output including the answer, reasoning, confidence, and cognitive classification (see Sections~\ref{sec:assessment} and \ref{sec:confidence-scoring} for details). The placeholder \texttt{\{expert\_role\}} is replaced with one of the definitions from Section A.1.

\begin{lstlisting}
You are analyzing a {question_type}-choice question.
{expert_role}

CRITICAL INSTRUCTION: You MUST provide your answer in the message.content field as a JSON object.
Your response MUST be concise and within 4000 tokens.

Special Handling:
- Questions may contain '[mask]' (replaced words) or '*' (replaced characters). Infer original content using context and options.
- For code (e.g., Java), prioritize syntax and logic analysis despite masks.
- If too ambiguous, choose the most likely answer and explain, or return "ERROR" with reasoning.

Please answer each question according to the following requirements:

1. Provide the Answer (Field name: "answerletter"):
   - Choose the correct option letter(s).
   - For {question_type}-choice questions, return {answer_format}.
   - Use uppercase letters.
   - Single-choice: return exactly ONE uppercase letter.
   - Multiple-choice: there are at least TWO correct options. Evaluate A-E individually and return ALL correct letters concatenated in alphabetical order without any separators (e.g., AB, BCD). Do NOT use commas, spaces, or other characters.

2. Provide Explanation (Field name: "reasoning"):
   - Keep explanation concise (max 100 words).
   - Focus on critical reasoning points.
   - Use bullet points or numbered lists.
   - Provide explanation in {output_language}.

3. Answer Confidence Rating (Field name: "confidence_answer_likert"):
   - Rate confidence (1-5): 1 = No confidence, 5 = Highly confident

4. Question Classification (Field name: "classification"):
   - Classify as "Lower-order" or "Higher-order"

5. Classification Confidence Rating (Field name: "confidence_classification_likert"):
   - Rate confidence in classification (1-5)

Output Requirements:
1. Output ONLY a JSON object in message.content
2. NO Markdown, code blocks, or quotes
3. All text in {output_language}
4. Keep explanation under 100 words
5. Use bullet points for clarity

Example JSON Format (Single-choice):
{
  "answerletter": "A",
  "reasoning": "Key point 1; Key point 2",
  "confidence_answer_likert": "5",
  "classification": "Higher-order",
  "confidence_classification_likert": "5"
}

Example JSON Format (Multiple-choice):
{
  "answerletter": "BD",
  "reasoning": "Reason 1; Reason 2",
  "confidence_answer_likert": "4",
  "classification": "Lower-order",
  "confidence_classification_likert": "4"
}
\end{lstlisting}

\textbf{Note:} The prompt enforces strict JSON output. Domain-specific expert roles from Section A.1 enhance reasoning accuracy.

\subsection{Translation Prompt}

The following prompt was used for cross-lingual translation of examination questions (English-to-Chinese and Chinese-to-English). Placeholders \texttt{\{source\_language\}}, \texttt{\{target\_language\}}, and \texttt{\{column\_names\}} are dynamically replaced.

\begin{lstlisting}
You are a professional {source_language} to {target_language} translation expert, specializing in computer science and technical documentation translation. Your task is to translate the {source_language} content provided by the user into {target_language}.

Please strictly follow the following requirements for translation:
1. Translate all {source_language} content into {target_language}
2. Maintain the original format and structure
3. Keep column names unchanged ({column_names})
4. Ensure translations are accurate, natural, and conform to {target_language} expression conventions
5. Use standard technical terminology for programming terms
6. Keep code sections unchanged, only translate text descriptions
7. Ensure translations are rigorous, professional, and preserve the original meaning
8. Do not return any answers, explanations, or comments
9. Do not add any explanatory text in parentheses
10. Perform pure translation only, do not provide any additional information
\end{lstlisting}

\textbf{Note:} \texttt{\{source\_language\}} and \texttt{\{target\_language\}} are set to "Chinese" or "English" based on automatic language detection. \texttt{\{column\_names\}} is replaced with actual column names from the dataset.

\subsection{Difficulty Rating Prompt}

The following prompt was used to evaluate question difficulty on a 5-point Likert scale. The placeholder \texttt{\{expert\_role\}} is replaced with a domain-specific definition from Section A.1.

\begin{lstlisting}
You are an expert exam analyst and {expert_role}

Your task: Evaluate ONLY the difficulty of the given multiple-choice question.

Rating scale (Likert 1-5):
Topic difficulty score
- Assign a numerical difficulty to the question based on the following criteria:
1= Very simple: Almost no thought required, relying only on the basics to answer;
2=Simple: involves simple concepts and logic, short thinking process;
3 = medium: requires some analysis and multi-step reasoning;
4 = Difficult: requires advanced logical reasoning and in-depth analysis;
5= Extremely difficult: Very complex, involving interdisciplinary knowledge and long time thinking.

Evaluation considerations:
- Required prior knowledge and concepts involved
- Reasoning depth, calculation or code tracing complexity
- Ambiguity and trickiness of distractors (options)
- For programming, consider API/semantics/edge-cases; for networking, consider protocols/topologies/config nuance; for office tasks, consider steps and feature combinations
- If content includes [mask] or * redactions, infer via context and options

Output requirements:
1. Output ONLY a JSON object in message.content
2. Use exactly one field: "difficulty_likert" with a value in {"1","2","3","4","5"}
3. No markdown, no extra fields, no explanation

Example:
{"difficulty_likert":"3"}

Your thinking process will be captured separately in reasoning_content.
\end{lstlisting}

\textbf{Note:} The prompt enforces strict JSON output for automated parsing. Domain-specific expert roles from Section A.1 improve evaluation accuracy.

\section{Supplementary Results: Perceived Difficulty Ratings}
\label{app:difficulty-ratings}

Table~\ref{tab:appendix-difficulty-ratings} presents the mean perceived difficulty ratings (on a five-point Likert scale, where 1 = very easy and 5 = very difficult) for each model across the six examination categories. Krippendorff's alpha is reported for inter-model agreement within each category and overall.

\begin{table}[!t]
    \centering
    \caption{Mean perceived difficulty ratings per model and examination category, with Krippendorff's $\alpha$ for inter-model agreement.}
    \label{tab:appendix-difficulty-ratings}
    \resizebox{\textwidth}{!}{
        \begin{tabular}{lcccccc}
            \toprule
            \multirow{2}{*}{\textbf{Model}} & \multicolumn{6}{c}{\textbf{Examination Category}} \\
            \cmidrule(lr){2-7}
            & CCNA & ICDL & Java & OCJP & MS Office & Chinese Net. Eng. \\
            \midrule
            GPT-5       & 1.42 & 1.04 & 1.27 & 2.17 & 1.17 & 1.30 \\
            DeepSeek-R1 & 1.11 & 1.10 & 1.76 & 1.99 & 1.51 & 1.55 \\
            Llama-3.3   & 2.36 & 1.66 & 2.00 & 2.58 & 1.91 & 2.21 \\
            Qwen-Plus   & 2.83 & 1.42 & 1.85 & 2.80 & 1.85 & 2.24 \\
            \midrule
            Krippendorff's $\alpha$ (per category) & -0.16 & -0.01 & 0.17 & 0.13 & 0.09 & 0.12 \\
            \bottomrule
            \multicolumn{7}{l}{\footnotesize Overall Krippendorff's $\alpha$ (across all categories): 0.13}
        \end{tabular}
    }
    \raggedright
    \footnotesize Note: Krippendorff's $\alpha$ values are rounded to two decimal places. Negative or near-zero values indicate low or no agreement (or even disagreement) among models in perceiving difficulty levels.
\end{table}

\section{Supplementary Results: Confidence-Accuracy Intervals}
\label{app:confidence-intervals}

Table~\ref{tab:appendix-confidence-intervals} provides detailed confidence-accuracy intervals by confidence level for each model.

\begin{table}[!t]
    \centering
    \caption{Confidence-accuracy intervals by confidence level for each model.}
    \label{tab:appendix-confidence-intervals}
    \resizebox{\textwidth}{!}{
        \begin{tabular}{lccccc}
            \toprule
            \textbf{Model} & \textbf{Confidence Level} & \textbf{Accuracy (\%)} & \textbf{CI Lower (\%)} & \textbf{CI Upper (\%)} & \textbf{Sample Size} \\
            \midrule
            GPT-5 & L1 & 0.0 & 0.0 & 0.0 & 1 \\
            GPT-5 & L2 & 25.0 & 0.0 & 50.0 & 8 \\
            GPT-5 & L3 & 42.1 & 21.1 & 63.2 & 19 \\
            GPT-5 & L4 & 70.2 & 60.6 & 78.9 & 104 \\
            GPT-5 & L5 & 94.2 & 92.7 & 95.7 & 936 \\
            DeepSeek-R1 & L2 & 37.5 & 12.5 & 75.0 & 8 \\
            DeepSeek-R1 & L3 & 71.4 & 42.9 & 100.0 & 7 \\
            DeepSeek-R1 & L4 & 43.9 & 29.3 & 58.5 & 41 \\
            DeepSeek-R1 & L5 & 92.0 & 90.4 & 93.6 & 1009 \\
            Llama-3.3 & L2 & 0.0 & 0.0 & 0.0 & 2 \\
            Llama-3.3 & L3 & 80.9 & 78.3 & 83.3 & 1008 \\
            Llama-3.3 & L4 & 50.0 & 0.0 & 100.0 & 2 \\
            Llama-3.3 & L5 & 87.5 & 78.6 & 94.6 & 56 \\
            Qwen-Plus & L2 & 50.0 & 16.7 & 83.3 & 6 \\
            Qwen-Plus & L3 & 20.0 & 7.5 & 32.5 & 40 \\
            Qwen-Plus & L4 & 59.2 & 44.9 & 73.5 & 49 \\
            Qwen-Plus & L5 & 92.4 & 90.5 & 93.9 & 973 \\
            \bottomrule
        \end{tabular}
    }
\end{table}

\end{document}